\documentclass[aps,pre,twocolumn,showpacs,superscriptaddress]{revtex4}

\usepackage{graphicx,subfigure}
\usepackage{amssymb}
\usepackage{amsmath}
\usepackage[british]{babel}
\usepackage{graphicx}
\usepackage{color}

\begin{document}

\title{Population splitting, trapping, and non-ergodicity in heterogeneous
diffusion processes}

\author{Andrey G. Cherstvy}
\email{a.cherstvy@gmail.com}
\affiliation{Institute for Physics \& Astronomy, University of Potsdam,
14476 Potsdam-Golm, Germany}
\author{Ralf Metzler}
\email{rmetzler@uni-potsdam.de}
\affiliation{Institute for Physics \& Astronomy, University of Potsdam,
14476 Potsdam-Golm, Germany}
\affiliation{Department of Physics, Tampere University of Technology, 33101
Tampere, Finland}
\affiliation{Mathematical Institute, University of Oxford, 24-29 St Giles',
Oxford OX1 3LB, United Kingdom}

\date{\today}

\begin{abstract}
We consider diffusion processes with a spatially varying diffusivity giving rise
to anomalous diffusion. Such heterogeneous diffusion processes are analysed for
the cases of exponential, power-law, and logarithmic dependencies of the
diffusion coefficient on the particle position. Combining analytical approaches
with stochastic simulations, we show that the functional form of the
space-dependent
diffusion coefficient and the initial conditions of the diffusing particles are
vital for their statistical and ergodic properties. In all three cases a weak
ergodicity breaking between the time and ensemble averaged mean squared
displacements is observed. We also demonstrate a population splitting of the
time averaged traces into fast and slow diffusers for the case of exponential
variation of the diffusivity as well as a particle trapping in the case of the
logarithmic diffusivity. Our analysis is complemented by the
quantitative study of the space coverage, the diffusive spreading of the
probability density, as well as the survival probability.
\end{abstract}


\maketitle

\section{Introduction} 

Anomalous diffusion of the power-law form \cite{bouchaud,report}
\begin{equation}
\label{msd}
\left<x^2(t)\right>\sim t^\beta
\end{equation}
of the mean squared displacement (MSD)
has been observed in a wide variety of systems. Depending on the value of the
anomalous diffusion exponent $\beta$ we distinguish subdiffusion ($0<\beta<1$)
and superdiffusion ($\beta>1$). The special cases are that of normal Brownian
motion ($\beta=1$) and wave-like, ballistic motion ($\beta=2$).

Examples for subdiffusion include the anomalous motion of charge carriers in
amorphous semiconductors \cite{scher}, the motion of tracer beads in polymer
melts \cite{amblard} and actin networks \cite{weitz}, the dynamics of sticky
particles along a surface \cite{chaikin}, or the spreading of tracer chemicals
in subsurface hydrology \cite{harvey}. Superdiffusion is observed in weakly
chaotic systems \cite{swinney}, in bulk-surface exchange controlled dynamics
in porous glasses \cite{stapf}, or for the motion of tracer beads in wormlike
micellar solutions \cite{ott}.

In particular, numerous cases of anomalous diffusion have been reported for the
motion of endogenous and artificial submicron tracers in living biological cells,
following substantial advances in single particle tracking and spectroscopic
tools over the last decade or so \cite{pt,saxton,franosch13,bress13}.
Thus, methods
such as video tracking, tracking by optical tweezers, or fluorescence correlation
spectroscopy have become routine tools to explore the motion of tracers such
as larger biomolecules or microbeads in vivo. The anomalous diffusion of
submicron-sized tracers is of interest for the understanding of biochemical
processes in the cell, but also offers insight
into the mechanical properties of the intracellular fluid and cellular mechanical
structures as the passive or active tracer motion represents the basis for
microrheology \cite{micro}.

Examples for in vivo subdiffusion include the motion of endogenous granules
(lipids or insulin) \cite{lene,tabei,taylor}, of fluorescently labelled RNA
molecules \cite{golding,weber}, of the tips (telomeres) of eukaryotic DNA and
loci of bacterial DNA \cite{bronstein,weber}, microbeads \cite{guigas,caspi},
viruses \cite{seisen01,brauchle02},
pigment organelles \cite{bruno09}, or of small proteins \cite{fradin05}.
Potassium channels resident in the plasma membranes of living cells were shown
to subdiffuse \cite{weigel}, as well as the motion of membrane proteins in the
Golgi membrane \cite{weiss03}. In simulations, subdiffusion of lipid and protein
molecules in bilayers and monolayers was observed
\cite{jeon-lipids,kneller,akimoto}. Superdiffusion in living cells is observed
for motor-driven transport of viruses \cite{seisen01}, microbeads \cite{caspi},
as well as magnetic endosomes \cite{robert}.

These experimental observations of anomalous diffusion have been modelled
theoretically in terms of different generalised stochastic processes
\cite{pt,igor,pccp,goychuk,saxton,franosch13,bress13}. The most popular
models include obstructed (coralled) diffusion \cite{saxton} that leads to a
turnover between free diffusion and a thermal plateau value. Transiently, this
process can be fitted with the law (\ref{msd}). Continuous time random walks
\cite{scher,montroll} are based on random walk processes, in which the pausing
time between successive jumps is power-law distributed such that no
characteristic time scale exists, leading to anomalous diffusion of the form
(\ref{msd}). In an external potential or in the presence of non-trivial boundary
conditions this continuous time random walk process is conveniently described in
terms of the fractional Fokker-Planck equation \cite{report,ffpe}. The resulting
motion of subdiffusive continuous time random walks in intrinsically noisy
environments was recently studied \cite{nctrw}. Fractional
Brownian motion \cite{mandelbrot} and the closely related fractional Langevin
equation \cite{lutz} are driven by Gaussian noise, which is long-ranged
correlated in time, again leading to behaviour (\ref{msd}). In the subdiffusive
regime these two correlated Gaussian processes are intimately connected with a
viscoelastic environment \cite{goychuk,weiss}.
Some of their properties are shared with scaled
Brownian motion \cite{fulinski}. The law (\ref{msd}) is also effected by the
geometrical constraints imposed to a particle diffusing on a support with a
fractal dimension \cite{havlin,klemm}. Superdiffusion is modelled in terms of
fractional Brownian motion or L{\'e}vy walks \cite{wong,zukla,lw1,aljaz}, a
class of continuous time random walks with spatiotemporal coupling.

Above theoretical approaches are based on the assumption that the environment
is homogeneous and isotropic, or that over the relevant time and length scales
of the measurement spatial variations of the environment in some sense are
averaged out. Yet there are clear indications that in biological cells the
environment effects strong variations of the local diffusion constant. Thus,
maps of the local cytoplasmic diffusion coefficient in bacterial \cite{elfDX11}
and eukaryotic \cite{langDX11} cells indeed demonstrate substantial spatial
variations. Significant changes of the diffusivity along the trajectory of
single tracer particles in cells may also be affected by transient binding as
well as the abundance of biochemical energy supply and transcription activity
in different compartments of eukaryotic nuclei \cite{platani}.

Descriptions in terms of space-dependent diffusion coefficients $D(x)$ are in
fact widely used
in hydrological applications to mesoscopically describe diffusion in heterogenous
porous media \cite{hagger95}. In particular, inhomogeneous versions of continuous
time random walk models for water permeation in porous ground layers were
developed recently \cite{hdp-ctrw}.

Mathematically, spatially and temporally varying diffusivities give rise to
anomalous sub- and superdiffusion in a range of stochastic models, compare
Refs.~\cite{srokowski06,srokowski08,silva11,fulinski,chechkin-inhom}. In
particular, Richardson type diffusion in turbulent media was modelled in terms
of heterogeneous diffusion processes (HDPs) \cite{maglom}. Power-law forms for
$D(x)$ were proposed to capture the diffusion of a particle on a fractal support
\cite{fractals-proc}, yet, as shown below, this approach gives rise to weakly
non-ergodic motion and is inherently different from
the ergodic motion on fractals \cite{igor,yazmin}. The weakly non-ergodic
properties of HDPs were studied recently \cite{fulinski,hdp13}.

Here we analyse in detail the motion of a diffusing particle subjected to a
space-dependent diffusion coefficient $D(x)$, for the cases of exponential,
power-law, and logarithmic $x$-dependencies. We demonstrate that these processes
effect anomalous diffusion of the form (\ref{msd}) of both sub- and superdiffusive
forms as well as an ultraslow, logarithmic time dependence. Moreover, we show
that despite their description in terms of a time local
diffusion equation, these processes exhibit a weak ergodicity breaking in the
sense that the time and ensemble averaged MSD do not
converge, even in the long time limit, see below. Our study reveals that the
dynamics of the diffusing particle may crucially depend on its initial position,
and that the time averaged MSD may exhibit a splitting of
the entire population of diffusing particles into faster and slower fractions.

In the following Section we briefly review the properties of weak ergodicity
violation of stochastic processes. Section \ref{model} introduces
the HDP process in detail. In Sections \ref{sec-power} to \ref{sec-log} we
investigate the power-law, exponential, and logarithmic dependence of $D(x)$.
Finally, in Section \ref{sec-out} we draw our conclusions and present a brief
outlook.

\section{Weak ergodicity breaking}

Commonly we characterise a stochastic process in terms of the ensemble averaged
MSD (\ref{msd}) defined through the spatial average of $x^2$,
\begin{equation}
\label{EAMSD}
\langle x^2(t)\rangle=\int x^2P(x,t)dx,
\end{equation}
over the probability density function (PDF) $P(x,t)$ to find the particle at
position $x$ at time $t$. An alternative way to calculate the MSD is via the
time average
\begin{equation}
\label{TAMSD}
\overline{\delta^2(\Delta)}=\frac{1}{T-\Delta}\int_0^{T-\Delta}\Big(x(t+\Delta)
-x(t)\Big)^2dt
\end{equation}
over the time series $x(t)$, whose length is $T$. In the time averaged MSD
$\overline{\delta^2(\Delta)}$ the differences of the particle positions as
separated by the lag time $\Delta$ are evaluated along the trajectory $x(t)$.
For a Brownian process, it can be shown that in the limit of long $T$ both
definitions of the MSD agree, $\langle x^2(\Delta)\rangle=\overline{\delta^2(
\Delta)}$ \cite{pt,pccp}, a manifestation of ergodicity in the Boltzmann sense.
Even when $T$ remains finite, a similar equivalence is obtained between the
ensemble averaged MSD (\ref{msd}) and the time averaged MSD $\overline{\delta^2
(\Delta)}$, once we additionally average over a sufficiently large number of
individual trajectories \cite{pt,pccp},
\begin{equation}
\label{EATAMSD}
\left<\overline{\delta^2(\Delta)}\right>=\frac{1}{N}\sum_{i=1}^N\overline{
\delta_i^2(\Delta)}.
\end{equation}

Once the process is non-stationary, the integral kernel $[x(t+\Delta)-x(t)]^2$
will depend on both $\Delta$ and $t$, and the equivalence between ensemble and
time averaged MSDs will break down, a phenomenon called weak ergodicity
breaking \cite{web}. In particular, subdiffusive continuous time random walk
processes exhibit the linear lag time dependence $\langle\overline{\delta^2(
\Delta)}\rangle\simeq\Delta$, contrasting the power-law form (\ref{msd}) of the
corresponding ensemble average \cite{he,pt,pccp,ariel}. Under confinement,
$\langle
x^2(t)\rangle$ converges to a plateau, whose value is defined in terms of the
second moment of the corresponding Boltzmann distribution, while the time
average scales with $\Delta$ as $\overline{\delta^2(\Delta)}\simeq
\Delta^{1-\beta}$ \cite{pt,pccp,pnas}. Concurrently, subdiffusive continuous
time random walk processes age in the sense that physical observables described
by this process explicitly depend on the time separation between initial system
preparation and start of the measurement \cite{johannes}. The linear scaling
of the time averaged MSD is also observed for correlated \cite{vincent} and
ageing \cite{lomholt} continuous time random walks, while their respective
ensemble averaged MSDs scale like Eq.~(\ref{msd}) or logarithmically in time.
Superdiffusive continuous time random walk processes of the L{\'e}vy walk type
exhibit an ultraweak violation of ergodicity in the sense that time and ensemble
averaged MSD only differ by a constant factor \cite{aljaz,zukla}.

Below we show a new variant of weak ergodicity breaking, namely, that under
certain initial conditions the time averaged MSD may scale like the square
root of the lag time, $\overline{\delta^2(\Delta)}\simeq\Delta^{1/2}$, while
the ensemble average exhibits the ultraslow scaling $\langle x^2(t)\rangle\simeq
\log^2(t)$.

Do all anomalous diffusion processes give rise to weakly ergodic behaviour? In
fact, there exists ergodic subdiffusive motion. One example is the motion on
a fractal support \cite{yazmin}. Another example is that of unbiased fractional
Brownian motion and the motion described by the fractional Langevin equation,
both reaching algebraically the ergodic behaviour \cite{deng,pccp}. However,
when a particle described by fractional Brownian or fractional Langevin equation
motion is confined, transiently non-ergodic behaviour is observed, and the
exponential relaxation to the thermal value of the ensemble averaged MSD
is replaced by an algebraically slow relaxation in the time averaged MSD
\cite{jae}.

How can different anomalous stochastic processes be identified based on recorded
single particle tracking data? During the recent years several complementary
methods have been presented
\cite{pt,igor,he,olivier,tejedor,yazmin,p-var,kepten,kevin,saxton,pccp}. The use
of multiple, complementary diagnosis tools simultaneously is of particular
importance. For
instance, when we analyse the velocity autocorrelation function, its shape
appears almost identical for fractional Brownian motion and confined
subdiffusive continuous time random walks \cite{pccp}. Among the applied
methods are the first passage behaviour \cite{olivier}, the mean maximal
excursion method \cite{tejedor}, analysis of the fractal dimension of the
trajectory \cite{tejedor,yazmin}, ratios of higher order moments \cite{tejedor},
the distribution function of amplitude scatter between different trajectories
\cite{he}, p-variation methods \cite{p-var}, and others \cite{kevin,saxton}.

\section{The HDP model and its analysis} 
\label{model}

We now turn to the HDP model for anomalous diffusion. We explicitly define the
process and briefly introduce the quantities used to analyse the special cases
for the spatial variation of the diffusion coefficient $D(x)$ investigated in
the following Sections, namely, power-law, exponential, and logarithmic
dependencies on $x$.

We start with the stochastic Langevin equation for the displacement $x(t)$
of a particle diffusing in the absence of an external potential in a medium
with the position-dependent diffusivity $D(x)$,
namely\begin{equation}
\label{langevin}
\frac{dx(t)}{dt}=\sqrt{2D(x)}~\zeta(t).
\end{equation}
Here, $\zeta(t)$ represents a Gaussian white ($\delta$-correlated) noise with
unit norm $\langle\zeta(t)\zeta(t')\rangle=\delta(t-t')$ and zero mean $\langle
\zeta(t)\rangle$=0. We interpret the nonlinear stochastic Eq.~(\ref{langevin})
with multiplicative noise in the Stratonovich sense \cite{risken}, both in our
theoretical analyses and in the simulations. After averaging over the noise
$\zeta(t)$, the diffusion equation for the PDF $P(x,t)$ has the symmetric form
\cite{hdp13}
\begin{equation}
\label{diff-equation-strat}
\frac{\partial P(x,t)}{\partial t}=\frac{\partial}{\partial x}\left[\sqrt{D(
x)}\frac{\partial}{\partial x}\left(\sqrt{D(x)}P(x,t)\right)\right].
\end{equation}

\begin{figure}
\includegraphics[width=9cm]{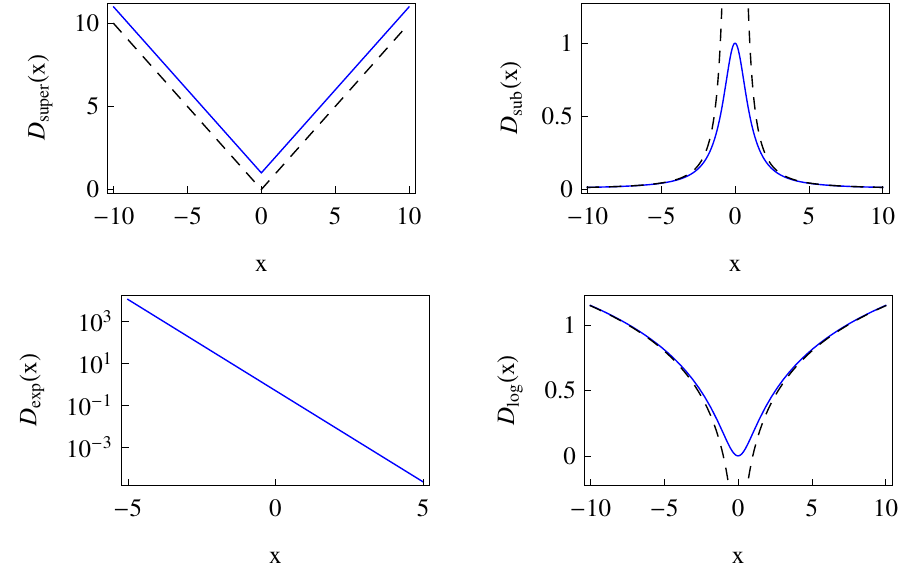}
\caption{Functional dependencies on the position variable $x$ of the diffusion
coefficients studied. The exact functional dependencies are represented by the
dashed lines, while the blue curves depict the regularised forms for $D(x)$
that were used in the simulations, see text.}
\label{fig-exp-log-poten}
\end{figure}

For this Markovian process with multiplicative noise, the different cases for
$D(x)$ we study in the following are depicted in Fig.~\ref{fig-exp-log-poten}.
Thus we consider the power-law shape
\begin{equation}
\label{power-diff}
D(x)=D_0|x|^{\alpha},
\end{equation} 
where the scaling exponent $\alpha$ may assume positive and negative values,
effecting sub- and superdiffusion, see below. While the form (\ref{power-diff})
turns out to be convenient for the analytical calculations, in the simulations
we employ regularised forms. Thus, for positive $\alpha$, the modified form
\begin{equation}
D_\text{super}=D_0(1+|x|^{\alpha})
\end{equation}
prevents the particle from getting stuck at the origin ($x=0$), while for
negative $\alpha$ the choice
\begin{equation}
D_\text{sub}=\frac{D_0}{1+|x|^{\alpha}}
\end{equation}
avoids the divergence of $D(x)$ at the origin. The power-law form
(\ref{power-diff}) along with the regularisations for sub- and superdiffusion
are shown in the top panel of Fig.~\ref{fig-exp-log-poten}.

In addition, we analyse the behaviour of the HDP for the exponential dependence
\begin{equation}
\label{D-exp}
D_{\mathrm{exp}}(x)=\frac{A^{2}}{2}e^{-2\alpha x},
\end{equation}
such that on the left semi-axis the diffusivity increases exponentially with
$|x|$, while on the positive semi-axis $D(x)$ decreases quickly. Finally, we
consider the logarithmic shape
\begin{equation}
\label{D-log}
D_{\mathrm{log}}(x)=\frac{A^{2}}{2}\frac{1}{2}\log\left[\left(\frac{x}{
\overline{x}}\right)^2+1\right],
\end{equation}
such that a trapping region of slow diffusion is created at small $x$ where
$D_{\mathrm{log}}(x)$ assumes a parabolic shape, while at $|x|\gg1$ the
diffusivity grows logarithmically like 
\begin{equation}
\label{log-diff-approx}
D_{\mathrm{log}}(x)\sim\frac{A^{2}}{2}\log\left[\frac{|x|}{\overline{x}}\right]. 
\end{equation}
In both cases the constants $A$ and $\alpha$ have the dimensions $\mathrm{cm}/
\mathrm{sec}^{1/2}$ and $1/\mathrm{cm}$, respectively, and we set $\overline{x}=
1$ below.
We assume that local thermal equilibrium is established on the length-scales
of spacial $D(x)$ variations. In Eq.~(\ref{D-log}) the addition of unity in
the logarithm prevents the divergence to minus infinity at the origin. The
exponential and logarithmic shapes for $D(x)$ are depicted in the bottom panel
of Fig.~\ref{fig-exp-log-poten}.

Numerically, following the Stratonovich interpretation the solution of
Eq.~(\ref{langevin}) requires an implicit mid-point iterative scheme for the
particle displacement $x_i$. At the simulation step $i+1$ we thus have
\begin{equation}
\label{langevin-simul}
x_{i+1}-x_i=\sqrt{2D([x_{i+1}+x_i]/2)}~(y_{i+1}-y_{i}),
\end{equation}
where  the increments of the Wiener process $(y_{i+1}-y_{i})$ represent a
centred, $\delta$-correlated Gaussian noise with unit variance. Unit time
intervals $\Delta t$ separate consecutive iteration steps in the simulations.
From a set of stochastic trajectories $x(t)$ generated for an initial particle
position $x(0)=x_0$, the ensemble and time averaged MSDs are computed. This
numerical scheme has recently been implemented for HDPs with a power-law
form \cite{hdp13}.

In what follows we evaluate the simulated time series $x(t)$ in terms of the
ensemble averaged MSD (\ref{EAMSD}), revealing different forms of sub- and
superdiffusion. To analyse the ergodic properties of the HDPs, the time
averaged MSD (\ref{TAMSD}) is evaluated along the trajectories as function of
the lag time $\Delta$. We also evaluate the additional average (\ref{EATAMSD})
over multiple trajectories.

For finite trajectories the time averaged MSD
(\ref{TAMSD}) between different trajectories will always vary. When the length
$T$ of the time series reaches very large values (ideally, it is taken to
infinity), the ergodicity breaking parameter \cite{he,rytov}
\begin{equation}
\label{EB1}
\mathrm{EB}=\lim_{T/\Delta\to\infty}\frac{\left<\left(\overline{\delta(
\Delta)^2}\right)^2\right>-\left<\overline{\delta(\Delta)^2}\right>^2}{\left<
\overline{\delta(\Delta)^2}
\right>^2}
\end{equation}
quantifies how reproducible individual realisations of the process are. At some
lag time $\Delta$, a vanishing ergodicity breaking parameter is a sufficient
condition for the ergodicity of a given stochastic process. A necessary condition
is that the ratio of the time and ensemble averaged MSDs is unity. As such a
ratio involves only the second moments, an additional ergodicity breaking
parameter can be defined as
\begin{equation}
\label{EB2}
\mathcal{EB}=\frac{\left<\overline{\delta^2(\Delta,T)}\right>}{
\left<x^2(\Delta)\right>}.
\end{equation}
Although this parameter is easier to compute analytically, it may strongly
depend on the initial conditions and is therefore not a universal feature
of a stochastic process.

The scatter distribution for the amplitude $\overline{\delta^2}$ of individual
trajectories around the mean $\langle\overline{\delta^2}\rangle$ is quantified
by the distribution
\begin{equation}
\phi(\xi)=\phi\left(\frac{\overline{\delta^2}}{\langle\overline{\delta^2}\rangle}
\right)
\end{equation}
in terms of the dimensionless variable $\xi$. It characterises the randomness of
individual time averaged MSDs
and yields additional information in how far the diffusion process deviates from
the ergodic behaviour.

For Brownian motion the finite-time scaling reads \cite{he} 
\begin{equation}
\label{eb-bm}
\mathrm{EB_{BM}}=\frac{4}{3}\frac{\Delta}{T}
\end{equation}
for the ergodicity breaking parameter, and 
\begin{equation}
\label{eb-bm1}
\phi_{\text{BM}}(\xi)\to\delta(\xi-1)
\end{equation}
for the amplitude scatter distribution at $T/\Delta\to\infty$. Both limiting
behaviours are in excellent agreement with simulations of Brownian motion (not
shown).

\section{Power-law varying diffusivity}
\label{sec-power}

\begin{figure}
\includegraphics[width=7cm]{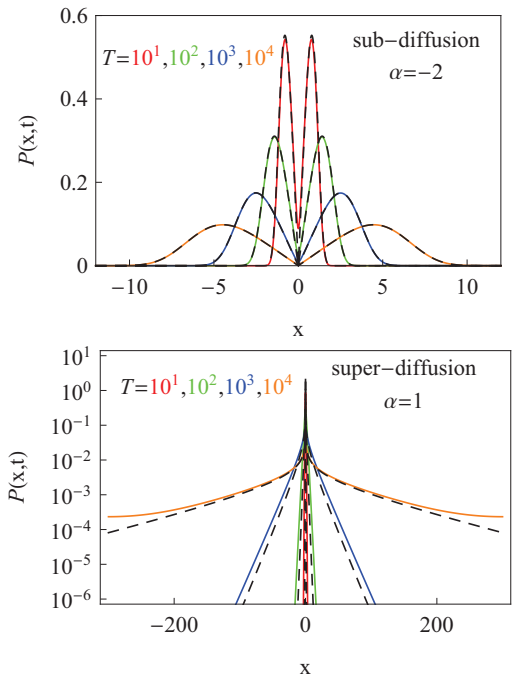}
\caption{The PDF for sub- and superdiffusive HDPs with power-law diffusivity
(\ref{power-diff}) and $D_0=1$. We show the analytical result (\ref{PDF-HDP})
for different trajectory lengths $T$ (coloured lines) and the numerical
solution of the dynamic
equation (\ref{diff-equation-strat}), represented by the dashed lines.}
\label{fig-power-pdf}
\end{figure}

Inserting the power-law form (\ref{power-diff}) of the diffusion coefficient
$D(x)$ into the diffusion equation (\ref{diff-equation-strat}), we recover
the PDF \cite{hdp13}
\begin{equation}
\label{PDF-HDP}
P(x,t)=\frac{|x|^{-\alpha/2}}{\sqrt{4\pi D_0t}}\exp\left(-\frac{|x|^{2-\alpha}}{
(2-\alpha)^{2}D_0t}\right)
\end{equation}
for the initial condition $P(x,0)=\delta(x)$. This equation, in turn, provides
the ensemble averaged MSD
\begin{equation}
\label{MSD-theory}
\langle x^{2}(t)\rangle=\Gamma\left(\frac{6-\alpha}{2(2-\alpha)}\right)
\frac{\left(2-\alpha\right)^{4/(2-\alpha)}}{\pi^{1/2}}(D_0t)^{2/(2-\alpha)}.
\end{equation}
According to Eq.~(\ref{MSD-theory}), for $\alpha<0$ the process is subdiffusive,
while superdiffusion emerges for $\alpha>0$. The limiting cases of Brownian
motion with $\langle x^2(t)\rangle=2D_0t$ corresponds to $\alpha=0$, and that
of ballistic motion for $\alpha=1$. The diffusion becomes increasingly fast
when $\alpha$ increases towards the limiting value 2. The PDF (\ref{PDF-HDP})
corresponds to a compressed Gaussian in the subdiffusive case ($\alpha<0$), i.e.,
we obtain
an exponential distribution in which the exponent of $x$ is larger than 2. In
the superdiffusive case ($0<\alpha<2$) the PDF (\ref{PDF-HDP}) becomes a
stretched Gaussian. Excellent agreement is observed between the theoretical
PDF (\ref{PDF-HDP}) and the numerical solution of the diffusion equation
(\ref{diff-equation-strat}), as demonstrated in Fig.~\ref{fig-power-pdf}.

Further analysis of the correlation function of consecutive increments of the HDP
process reveals the anti-persistent nature for the subdiffusion case, while
persistent correlations accompany the superdiffusive case \cite{hdp13}. The
analytical result for the velocity-velocity correlation function can be shown
to resemble the correlation function of fractional Brownian motion
\cite{hdp13}. 

The trajectory-to-trajectory averaged time averaged MSD (\ref{EATAMSD}) of the
HDP process with power-law form (\ref{power-diff}) of the diffusion coefficient
takes on a linear dependence on the lag time $\Delta$ \cite{hdp13},
\begin{eqnarray}
\nonumber
\left<\overline{\delta^2(\Delta)}\right>&=&\Gamma\left(\frac{6-\alpha}{2(2-
\alpha)}\right)\frac{(2-\alpha)^{4/(2-\alpha)}}{\pi^{1/2}}\\
&&\times D_0^{2/(2-\alpha)}\Delta T^{\alpha/(2-\alpha)}.
\label{TAMSD-theory}
\end{eqnarray}
This result can be rewritten in the form
\begin{equation}
\label{ea-tamsd}
\left<\overline{\delta^2(\Delta)}\right>=\left<x^2(\Delta)\right>\left(\frac{
\Delta}{T}\right)^{-\alpha/(2-\alpha)},
\end{equation}
introducing the strong ageing dependence on the measurement time $T$: as function
of the lag time the motion slows down. We can alternatively express this
statement in the form $\langle\overline{\delta^2(\Delta)}\rangle\simeq D_{\text{
eff}}(T)\Delta$, such that this effective diffusion coefficient has the scaling
\begin{equation}
D_{\text{eff}}(T)\simeq T^\frac{\alpha}{2-\alpha}.
\end{equation}
The functional relation (\ref{ea-tamsd}) between ensemble
and time averaged MSDs is identical to the one observed for subdiffusive
continuous time random walk processes \cite{he} as well as continuous
time random walk processes with correlated waiting times \cite{vincent}.

The scatter distribution $\phi\left(\overline{\delta^2(\Delta)}/\langle\overline{
\delta^2(\Delta)}\rangle\right)$ in the sub- and superdiffusive cases,
respectively, follows a Rayleigh-like and a generalised Gamma distribution
\cite{hdp13}. Moreover for a fixed length $T$ of the underlying time
series $x(t)$, the scatter distribution $\phi(\xi)$ stays nearly constant for
varying lag times $\Delta$. In other words, the degree of fluctuations around
the mean $\langle\overline{\delta^2(\Delta)}\rangle$ is approximately
invariant along the HDP trajectories.

For subdiffusion with $\alpha<0$ we see from Eq.~(\ref{ea-tamsd}) that the
time averaged MSD is much smaller than the ensemble averaged MSD,
$\langle\overline{\delta^2(\Delta)}\rangle\ll\left< x^2(\Delta)\right>$,
as long as $\Delta\ll T$. In contrast $\langle\overline{\delta^2(\Delta)}\rangle
\gg\left< x^2(\Delta)\right>$ for superdiffusion with $\alpha>1 $.
Because of the larger amplitude spread quantified by the scatter distribution
$\phi$ and its strongly asymmetric shape, the EB parameter for the case of
superdiffusion is systematically larger than the one for subdiffusion:
$\mathrm{EB}_{\text{super}}\approx1.4$ compared to $\mathrm{EB}_\text{sub}
\approx0.4$ for $\alpha=-2$ and $\alpha=1$, respectively. This observation as
well as the $\Delta$-dependence of the second EB parameter
\begin{equation}
\mathcal{EB}=\left(\frac{\Delta}{T}\right)^{-\alpha/(2-\alpha)}
\end{equation}
are supported by computer simulations performed according to the Stratonovich
scheme (not shown).

\section{Exponentially varying diffusivity}
\label{sec-exp}

We now turn to the exponentially varying diffusion coefficient (\ref{D-exp}).
We characterise the stochastic properties of this process with the same
quantities studied above, i.e., the PDF, the time and ensemble averaged MSDs,
the scatter distribution, and the ergodicity breaking parameters. In addition,
we explore the initial position-induced population splitting into fast and slow
walkers, the diffusion fronts, and the effective exploration of space.

Exponential distributions of the diffusion coefficient have been used to describe
the motion dynamics of parasitic nematodes \cite{nematodes}, or for the
irradiation-enhanced diffusion of impurities where the exponential variation is
effected by the decay of the radiation when it penetrates an absorbing medium
\cite{irradiate}. Finally, an exponential rate of morphogen degradation was
applied in a reaction-subdiffusion model for cell development \cite{santos}.

\subsection{PDF and ensemble averaged MSD}

\begin{figure}
\includegraphics[width=8cm]{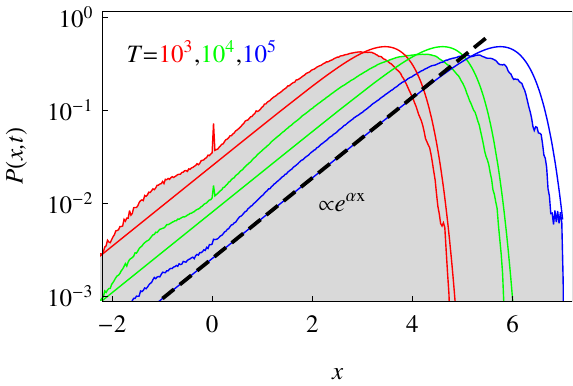}
\caption{The universal PDF shape for HDPs with $D_\mathrm{exp}(x)$, obtained
from simulations with initial condition $x_0=0$. The results are compared with
the theoretical result (\ref{exp-pdf}). The trace lengths $T$ are indicated, and
we chose $\alpha=1$ and $A=1$. The data were averaged over $N=400$ trajectories.
After averaging, the PDF for short traces still contains a small spike at the
initial position.}
\label{fig-exp-pdf-zero}
\end{figure}

To obtain the PDF for the HDP with the exponential $x$-dependence (\ref{D-exp}),
following the same steps as for the power-law form for $D(x)$ analysed in
Ref.~\cite{hdp13}, we employ the standard transformation of variables
\cite{srokowski09}
\begin{equation}
y(x)=\int^x\frac{dx'}{\sqrt{2D_{\text{exp}}(x')}}=\frac{\exp(\alpha x)}{
\alpha A}.
\label{var-change}
\end{equation}
Here $y(t)$ in the Stratonovich sense corresponds to the Wiener process, whose
PDF is the standard Gaussian
\begin{equation}
p(y,t)=\frac{1}{\sqrt{2\pi t}}\exp\left(-\frac{y^2}{2t}\right),
\label{wiener-pdf}
\end{equation}
Together with the normalisation condition $\int_{-\infty}^{\infty}P(x,t)dx=1$,
and the probability conservation law, from Eq.~(\ref{wiener-pdf}) the
normalised PDF of the HDP with exponentially varying diffusivity assumes the
unimodal double-exponential form
\begin{equation}
P(x,t)=\frac{2}{A}\frac{\exp(\alpha x)}{\sqrt{2\pi t}}\exp\left(-\frac{1}{
2t}\left[\frac{e^{\alpha x}-e^{\alpha x_0}}{A\alpha}\right]^2\right),
\label{exp-pdf-x0}
\end{equation}
for arbitrary initial value $x_0$. In the limit $x\to-\infty$ the PDF of the
particle after time $t$ features the exponential tail,
\begin{eqnarray}
\nonumber
P(x,t)&\sim&\frac{2}{A}\frac{\exp(\alpha x)}{\sqrt{2\pi t}}\exp\left(-\frac{\exp(
2\alpha x_0)}{2tA^2\alpha^2}\right)\\
&\sim&\frac{2}{A}\frac{\exp(\alpha x)}{\sqrt{2\pi t}},
\label{exp-pdf-exp-tail}
\end{eqnarray}
where in the second approximation we also took the long time limit. At large
values of $x$ the PDF decays sharply in a double-exponential fashion.

\begin{figure}
\includegraphics[height=8cm,angle=270]{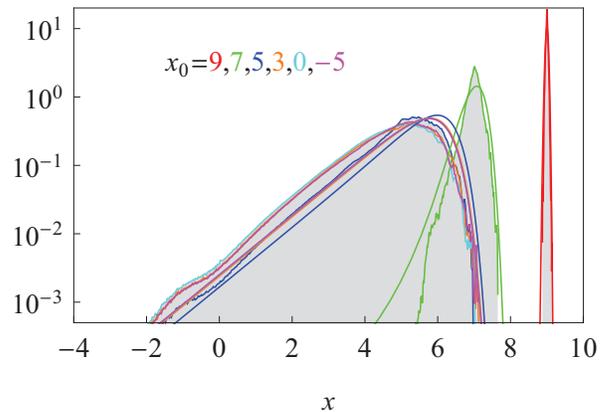}
\caption{PDF of the HDP with $D=D_{\mathrm{exp}}(x)$ for various initial
positions $x_0$ of the process. The smooth curves represent the theoretical
result, Eq.~(\ref{exp-pdf-x0}). The parameters were chosen as $T=10^5$,
$\alpha=1$, and $A=1$, and $N=400$ traces were analysed for each of the shown
profiles.}
\label{fig-exp-pdf-x0}
\end{figure}

For further analysis we assume that the initial condition has a sufficiently
large modulus on the left semi-axis, that is, $|x_0|\ll(2\alpha)^{-1}$. In
this case the PDF becomes
\begin{equation}
P(x,t)=\frac{2}{A}\frac{\exp(\alpha x)}{\sqrt{2\pi t}}\exp\left(-\frac{\exp(
2\alpha x)}{2t A^2\alpha^2}\right)
\label{exp-pdf}
\end{equation}
Its maximum is located at
\begin{equation}
\label{xmax}
x_{\text{max}}=\frac{\log(\alpha^{2} A^2 t)}{2\alpha},
\end{equation}
where the PDF has the value
\begin{equation}
P(x_{\text{max}})=\sqrt{\frac{2\alpha^2}{\pi e}}.
\end{equation}
Interestingly, the temporal shift of of the maximum position is logarithmic in
time, while the value of the PDF at this maximum remains constant. We compare
the functional forms (\ref{exp-pdf}) of the PDF with simulations results in
Fig.~\ref{fig-exp-pdf-zero}, observing very favourable agreement. For the
non-zero initial position $x_0$, the PDF is shown in Fig.~\ref{fig-exp-pdf-x0},
also exhibiting good agreement with the analytical form (\ref{exp-pdf-x0}).

The MSD may now be obtained from the PDF (\ref{exp-pdf}) simply by integration.
The exact result reads
\begin{equation}
\label{MSD-exp-theory}
\langle x^2(t)\rangle=\frac{1}{4\alpha^2}\left(A_1+A_2\log\Big[\alpha^2A^2t\Big]
+\log^2\Big[\alpha^2A^2t\Big]\right),
\end{equation}
where $\gamma\approx0.57721$ is the Euler-Mascheroni constant (or Euler's
constant), and we also define the two abbreviations
\begin{equation}
A_1=\pi^2/2+\log^2[2]+\gamma^2+2\gamma\log[2] \approx 6.55
\end{equation}
and
\begin{equation}
A_2=2\gamma+2\log[2]\approx2.54,
\end{equation}
Thus, at long times $t\gg(\alpha^2 A^2)^{-1}$ we thus observe the logarithmic
behaviour
\begin{equation}
\langle x^2(t)\rangle\sim\frac{1}{4\alpha^2}\log^2\Big[\alpha^2A^2t\Big].
\end{equation}
Formula (\ref{MSD-exp-theory}) could also be obtained directly from the
stochastic equation (\ref{langevin}) in the following way. With the transformation
(\ref{var-change}) and the distribution (\ref{wiener-pdf}) of the Wiener process,
the MSD (\ref{MSD-exp-theory}) results from the averaging $\int_{-\infty}^{
\infty}p(y,t)x^2(y)dy$. We note that for general initial condition $x_0$ we
could not find an analytical result for the MSD. Numerical analysis confirms
that the MSD shows the logarithmic time dependence (\ref{MSD-exp-theory}), and
in the long time limit exactly converges to this form.

The logarithmic scaling of the ensemble averaged MSD (\ref{MSD-exp-theory})
resembles that of other ultraslow processes. Thus, continuous time random walks
with logarithmic distribution of waiting times exhibit a slow logarithmic
growth of the MSD \cite{katja} as well as ageing continuous time random
walks \cite{lomholt}. The most prominent example for logarithmic time evolution
is that of Sinai diffusion, the motion of a random walker in a random force
field, where the ensemble averaged MSD follows the law $\langle x^2_\text{Sinai}(
t)\rangle\sim\log^4t$ \cite{sinaj,doussal-sinai}. Remarkably, our PDF
(\ref{exp-pdf}) is identical to that in the Sinai model with ageing in the
limit when the height of the barriers for consecutive jumps of a particle vary
linearly with position. This leads to an exponential dependence of the effective
diffusion coefficient and also to a $\langle x^{2}(t)\rangle \sim\log^2t$
scaling for the ensemble averaged MSD \cite{doussal98}.

To further quantify the dynamics of the diffusing particles, we performed
stochastic simulations according to the scheme (\ref{langevin-simul}). From
the generated trajectories $x(t)$ of the walker the PDFs were computed for
different starting positions $x_0$ and trace lengths $T$. For negative $x_0$ the
particles start in the domain of fast diffusion [large $D(x)$, compare
Fig.~\ref{fig-exp-log-poten}] and rapidly escape the negative
semi-axis. Typically, they become trapped on the positive semi-axis, where
$D(x)$ is smaller. For large positive initial position, $x_0\gg1$, the PDF is
sharply peaked as the particles on average remain trapped in the region of
extremely low (exponentially small) diffusivity. This peak slowly spreads for
longer traces. 

When $x_0$ becomes smaller than some `critical' value, the PDF follows a
universal asymmetric shape with an exponential tail at $x<0$. On the positive
semi-axis, a sharp double-exponential drop-off of the PDF is observed, with a
$T$-dependent location. These trends are in agreement with Eqs.~(\ref{exp-pdf})
and (\ref{exp-pdf-exp-tail}), whose functinal form is compared with
the simulations results in
Fig.~\ref{fig-exp-pdf-x0}. For longer $T$, the maximum of the PDF shifts to
larger $x$ values, in agreement with the theoretical prediction (\ref{xmax}),
compare Fig.~\ref{fig-exp-pdf-zero}.
The ensemble averaged MSD obtained from the generated trajectories closely
follows the $\sim\log^2[t]$ asymptote given by Eq.~(\ref{MSD-exp-theory}). At
$x_0\gg1$ the ensemble averaged MSD relaxes to this asymptote at later times
because the
particles are initially trapped in the exponentially slow diffusion region,
resulting in a $x_0^2$-plateau at short times, see Fig.~\ref{fig-exp-tamsd}.

\subsection{Time averaged MSD}

\begin{figure}
\includegraphics[width=7cm]{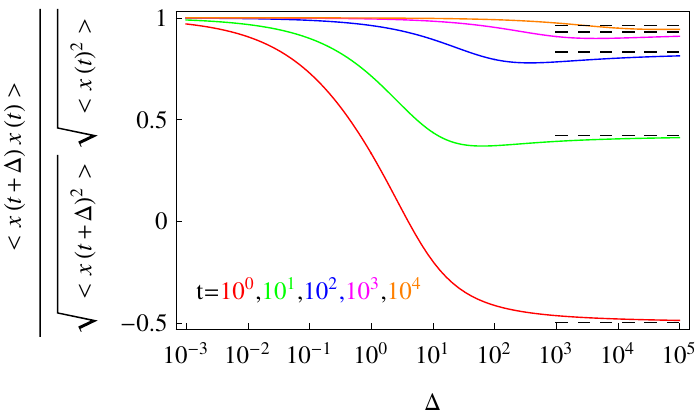}
\caption{The normalised position correlations (\ref{exp-pi-y}) for exponentially
distributed diffusion coefficient. Parameters: $\alpha=1$ and $A=1$. The dashed
lines indicates Eq.~(\ref{asterisk}).}
\label{fig-pos-cor-exp}
\end{figure}

To calculate the time averaged MSD we need to obtain the position
auto-correlation function, $\langle x(t_1)x(t_2)\rangle$. Using the two-point
probability density function for the Wiener process (without loss of generality,
$t_2>t_1$),
\begin{equation}
\pi(y_2,t_2|y_1,t_{1})=\frac{1}{\sqrt{2\pi(t_2-t_1)}}\exp\left(-\frac{(y_2
-y_1)^2}{
2(t_2-t_1)}\right),
\label{pi-wiener}
\end{equation}
one obtains for the positional correlation that
\begin{eqnarray}
\nonumber
\left\langle x(t)x(t+\Delta)\right\rangle&=&\frac{1}{4\alpha^2}\int_{-\infty}^{
\infty}dy_1\int_{-\infty}^{\infty}dy_2\\[0.2cm]
\nonumber
&&\hspace*{-1.8cm}
\times\log[(\alpha Ay_1)^2]\log[(\alpha Ay_2)^2]\\[0.2cm]
&&\hspace*{-1.8cm}
\times\pi(y_2, t+\Delta|y_1,t)p(y_1,t),
\label{exp-pi-y}
\end{eqnarray}
where we again use the trick of choosing a sufficiently negative initial
condition for convenience. After integration we arrive at ($\Delta<t$)
\begin{eqnarray}
\nonumber
\left< x(t)x(t+\Delta)\right>=\frac{1}{4\alpha^2}\left\{A_1-\frac{1}{2}\arctan
^2\left[\frac{2\sqrt{\Delta t}}{t-\Delta}\right]\right\}\\[0.2cm]
\nonumber
&&\hspace*{-7.6cm}
-\text{4~arccot}[\sqrt{\Delta/t}] \arctan[\sqrt{\Delta/t}]\\[0.2cm]
\nonumber
&&\hspace*{-7.6cm}
+\frac{A_2}{2}(\log[\alpha^2A^2t]+\log[\alpha^2A^2(t+\Delta)])\\[0.2cm]
&&\hspace*{-7.6cm}
+\log[\alpha^2A^2t] \log[\alpha^2A^2(t+\Delta)].
\label{exp-corr}
\end{eqnarray}
For $\Delta=0$ this expression coincides with the regular ensemble averaged
MSD (\ref{MSD-exp-theory}), as it should. The functional dependence of the
positional correlations is shown in Fig.~\ref{fig-pos-cor-exp}. In the limit
$\Delta\gg t$ the position autocorrelation function (\ref{exp-corr}) approaches
\begin{equation}
\label{asterisk}
\left< x(t)x(t+\Delta)\right>\sim\frac{\frac{1}{2}A_2+\log\left(\alpha^2 A^2t
\right)}{\left[A_1+A_2\log\left(\alpha^2 A^2t\right)+\log^2\left(\alpha^2 A^2t
\right)\right]^{1/2}}.
\end{equation}

\begin{figure}
\includegraphics[width=9cm]{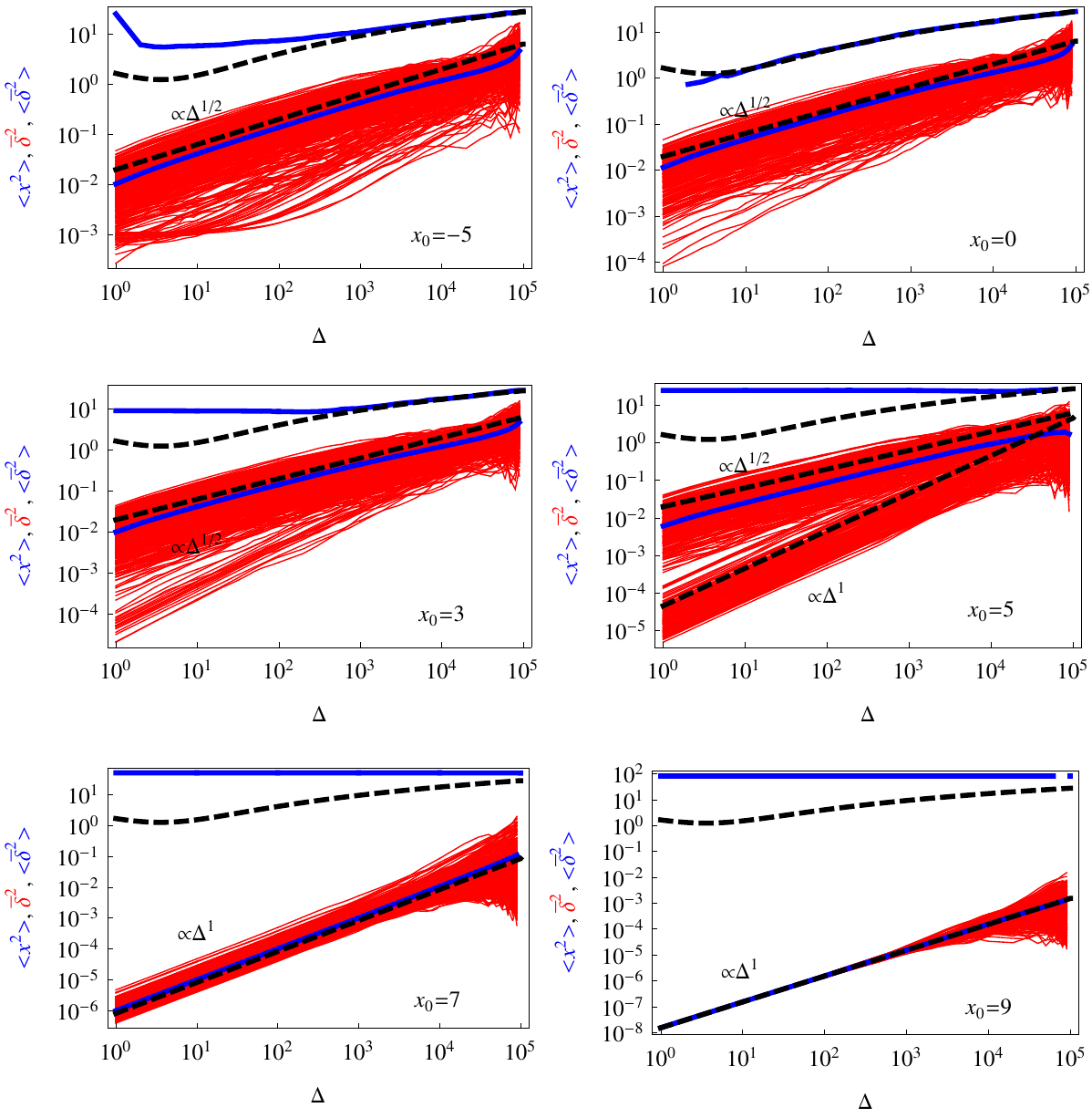}
\caption{Ensemble averaged MSD $\langle x^{2}(t)\rangle$  (upper blue curves in
each panel), time average MSD $\overline{\delta^2}$ of individual trajectories
(red curves), and mean time averaged MSD $\langle\overline{\delta^2}\rangle$
(lower blue curves in each panel). The starting positions $x_0$ for each panel
are indicated. The dashed black curves represent Eq.~(\ref{MSD-exp-theory})
for the ensemble averaged MSD and Eqs.~(\ref{exp-simple-TAMSD}) and
(\ref{tamsd-exp-effective-d}) for the two populations, respectively. The
$\overline{\delta^2}$ are shown with log-sampled points along the
$\Delta$-axis. At $x_0$=5 a splitting of $\overline{\delta^2}$ into a slow
($\overline{\delta^2}\simeq\Delta^{1/2}$) and fast ($\overline{\delta^2}\simeq
\Delta$) fraction is most pronounced. Parameters: $T=10^5$, $\alpha=1$, $A=1$.
$N=400$ trajectories were simulated to produce the trajectory-average.}
\label{fig-exp-tamsd}
\end{figure}

For the time averaged MSD (\ref{TAMSD}) a simple scaling argument can be
established in the limit of short lag times, $\Delta\ll T$. To this end, we
notice that the time averaged MSD
\begin{eqnarray}
\nonumber
\left<\overline{\delta^2(\Delta)}\right>
&=&\frac{1}{T-\Delta}\int\limits_{0}^{T-\Delta}\Big[\left<x^2(t+\Delta)\right>\\
&&+\left<x^2(t)\right>-2\left<x(t+\Delta)x(t)\right>\Big]dt
\end{eqnarray}
contains three correlators in the integrand. Expanding both the MSD
(\ref{MSD-exp-theory}) and the two-point correlator (\ref{exp-corr})
in $\Delta$, in the limit $\Delta\ll T$ we find that
\begin{equation}
\label{exp-simple-TAMSD}
\left<\overline{\delta^2(\Delta)}\right>\sim\frac{1}{T-\Delta}\int\limits_0^{
T-\Delta}\frac{\pi}{\alpha^2}\sqrt{\frac{\Delta}{t}}dt\approx\frac{2\pi}{
\alpha^2}\left({\frac{\Delta}{T}}\right)^{1/2}.
\end{equation}
The square-root scaling $\langle\overline{\delta^2(\Delta)}\rangle\simeq\Delta^
{1/2}$ is very distinct from the linear scaling observed for subdiffusive
continuous time random walk processes \cite{pt,pccp,he} as well as for time
correlated continuous time random walks \cite{vincent}, for ageing
continuous time random walks \cite{lomholt}, and for HDPs with power-law
distributed diffusivities \cite{hdp13} presented in the previous Section.
For initial conditions $x_0$ that are far away from zero on the negative
semi-axis, i.e., particles starting in the high-diffusivity region, the
approximate scaling (\ref{exp-simple-TAMSD}) agrees pretty nicely with
simulations results, as shown in Fig.~\ref{fig-exp-tamsd}.   

In the opposite case of large positive $x_0$, the integration of
Eq.~(\ref{langevin}) yields (at $\alpha x_0\ll1$)
\begin{equation}
\label{exp-x-via-y-large-x0}
x(y)\approx x_0+e^{-\alpha x_0}Ay(t),
\end{equation}
and after elementary averaging we find
\begin{equation}
\label{tamsd-exp-effective-d}
\left<\overline{\delta^2(\Delta)}\right>\approx2D_\text{exp}(x_0)\Delta.
\end{equation} 
Therefore, the time averaged MSD in this limit displays the linear behaviour,
that we observe for both Brownian processes as well as the above mentioned
anomalous diffusion processes. The effective diffusivity naturally depends on the
initial position $x_0$ of the particle. Eqs.~(\ref{exp-simple-TAMSD}) and
(\ref{tamsd-exp-effective-d}) reveal the exponents $\frac{1}{2}$ and $1$ of the
time averaged MSDs for these two extreme choices of $x_0$ that appear clearly
distinguished in Fig.~\ref{fig-exp-tamsd}. We note already here that when the
initial position $x_0$ is shifted towards more positive values, individual
trajectories become more reproducible (Fig.~\ref{fig-exp-tamsd}), as detailed
more quantitatively now.
 
\subsection{Amplitude scatter and ergodicity breaking parameter}

\begin{figure}
\includegraphics[width=9cm]{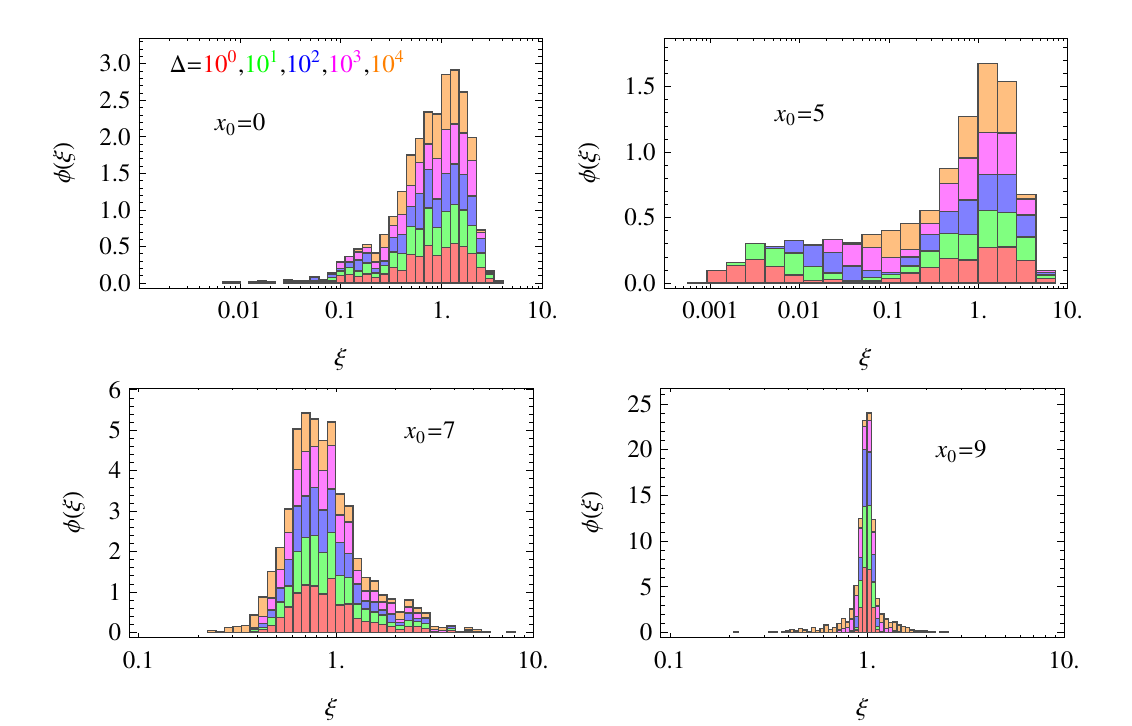}
\caption{Amplitude scatter distribution $\phi$ of individual time averaged MSDs
$\overline{\delta^2}$ for different initial positions $x_0$ for HDPs with
exponentially distributed diffusion coefficient. In each panel, the colours of
the stacked histograms correspond to different lag times $\Delta$ along the
$\overline{\delta^2(\Delta)}$ traces (the bins do not overlap). In the $x_0=5$
panel, the $\overline{\delta^2}$ traces at short lag times $\Delta$ clearly
split into two sub-populations (red bars), while at large $\Delta$ these two
distributions merge (orange bars). The parameters are the same as in
Fig.~\ref{fig-exp-tamsd}.}
\label{fig-phi-xi}
\end{figure}

\begin{figure}
\includegraphics[width=8cm]{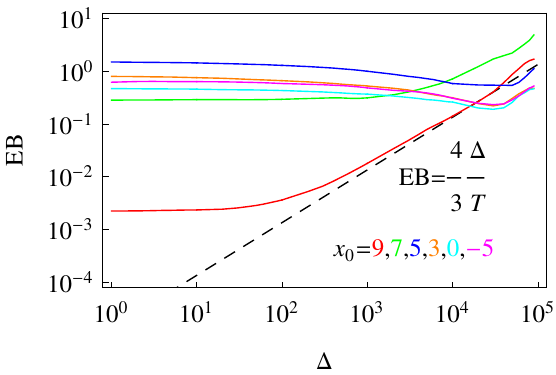}
\caption{Ergodicity breaking parameter as a function of the lag time for
varying initial particle positions $x_0$ for HDPs with exponentially
distributed diffusion coefficient. The parameters are the same as in
Fig.~\ref{fig-exp-tamsd}.}
\label{fig-exp-eb}
\end{figure}

As shown in Figs.~\ref{fig-exp-tamsd} and \ref{fig-phi-xi}, the time averaged MSD
exhibits a pronounced
amplitude scatter. This effect becomes increasingly stronger when the initial
position is more negative, i.e., when the particle is initially placed in the
high diffusivity region. For increasingly positive initial position the scatter
of individual $\overline{\delta^2(\Delta)}$ is reduced, in the panel for $x_0=9$
the trajectories are almost perfectly reproducible for shorter lag times.
Generally, ensemble and time averaged MSDs do not coincide
(Fig.~\ref{fig-exp-tamsd}), a feature that
clearly indicates a weak ergodicity breaking. This is further detailed in terms
of the ergodicity breaking parameter in Fig.~\ref{fig-exp-eb}. The non-ergodic
behaviour is due to the strong non-uniformity of the environment over typical
length scales of the diffusive motion. The scaling of the time averaged MSD
follows Eqs.~(\ref{exp-simple-TAMSD}) and (\ref{tamsd-exp-effective-d}) for
negative and positive values of the initial positions $x_0$ with large modulus
$|x_0|$, respectively (Fig.~\ref{fig-exp-tamsd}).
For smaller modulus of the initial position $x_0$, the
amplitude scatter of individual traces $\overline{\delta^2}$ becomes reduced at
longer lag times $\Delta$, i.e., the width of the scatter distribution $\phi$
for the longer $\Delta$ decreases,
as confirmed in Fig.~\ref{fig-phi-xi}. This phenomenon
is due to the fact that when the initial condition is further left on the axis
the PDFs tend to converge (Fig.~\ref{fig-exp-pdf-x0}). Thus at later stages
of the trajectories the particles' most probable location increasingly localises,
thus effecting smaller scatter between individual amplitudes, i.e., smaller
differences in the particle positions.

When the particle initial position is in
a slow-diffusion region, $x_0\gg1$, the HDP turns nearly ergodic and the amplitude
scatter increases when $\Delta$ becomes comparable to the overall length $T$ of
the time series. Using expression
(\ref{exp-x-via-y-large-x0}), one can show that the ergodicity breaking
parameter (\ref{EB1}) in the limit $\Delta\ll T$ vanishes to first order as 
\begin{equation}
\label{eb-0}
\text{EB}_{\text{exp}}\sim\frac{4}{3}\frac{\Delta}{T},
\end{equation} 
in agreement with computer simulations, which coincides with the result for
regular Brownian motion, Eq.~(\ref{eb-bm}). Note, however, that despite the
lack of amplitude scatter and the Brownian-style behaviour of the ergodicity
breaking parameter, this process remains weakly non-ergodic due to the
disparity between ensemble and time averaged MSDs.
Note that for $\Delta/T\ll1$ the HDP approaches the ergodicity differently
depending on the trace length $T$. Namely, for nearly ergodic starting
positions $x_0 \gg 1$, for shorter $T$ the $\textrm{EB}$ value is much closer to
the Brownian asymptote, compare Fig. \ref{fig-exp-eb} and Fig. A1 in the
Appendix. 

\subsection{Population splitting and exploration of space}

\begin{figure}
\includegraphics[width=7cm]{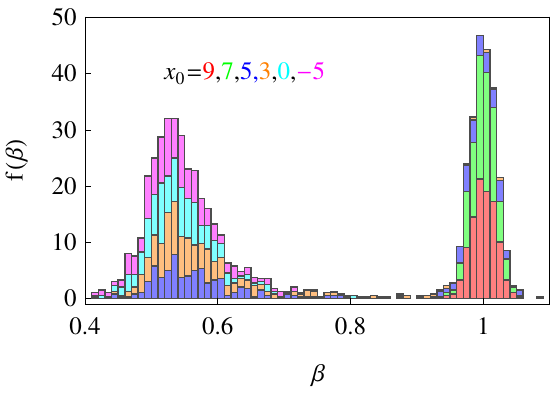}
\caption{Evolution of the scaling exponent $\beta$ of the time averaged MSD
$\overline{\delta^2}$ at short times for different initial positions $x_0$
for HDPs with exponentially varying diffusivity. A clear population splitting
is observed for smaller $x_0=5$, with maxima centred around the
predicted values $\frac{1}{2}$ and $2$ (blue bars). The parameters are the same
as in Fig.~\ref{fig-exp-tamsd}.}
\label{fig-exp-split}
\end{figure}

\begin{figure}
\includegraphics[width=7cm]{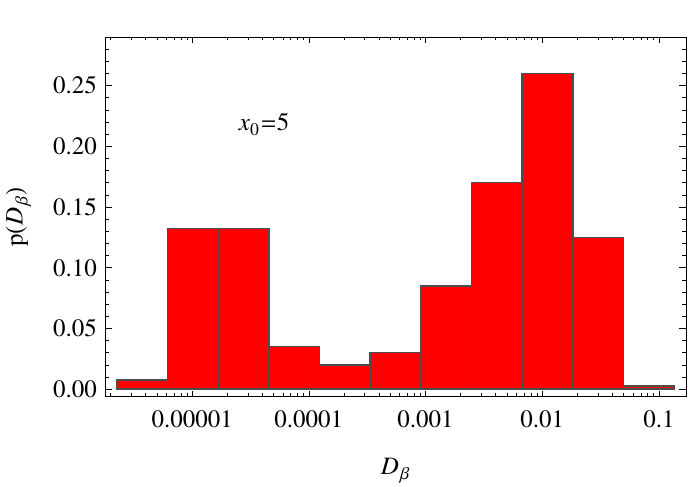}
\caption{Population splitting of apparent diffusion coefficients for HDPs with
exponentially varying diffusion coefficient, computed for the same parameters
as in Fig.~\ref{fig-exp-tamsd}.}
\label{fig-exp-D-split}
\end{figure}

Computer simulations show that at intermediate $x_0$ a population splitting
occurs between a slow fraction following the square-root scaling of the
time averaged MSD,
\begin{equation}
\overline{\delta^2(\Delta)}\simeq\Delta^{1/2}
\end{equation}
and an apparently ergodic fraction with the standard linear scaling
$\overline{\delta^2}\sim\Delta$. This is one of the the main features of the
$\overline{\delta^2}$ traces for the case of exponential variation of the
diffusion coefficient as function of the particle position $x$. Such a two-phase
dynamics is observed due to the fast particles starting at $x<0$ and the nearly
ergodic, slow walkers starting at $x\gg1$. With increase of $x_0$ the scaling
exponent $\beta$ for the initial region $\Delta\gg T$ of the trajectories,
\begin{equation}
\overline{\delta^2(\Delta\gg T)}\simeq D_{\beta}\Delta^\beta,
\end{equation}
changes from $\beta=$1/2 to $\beta$=1, as predicted by
Eqs.~(\ref{exp-simple-TAMSD}) and (\ref{tamsd-exp-effective-d}) \cite{beta-note},
splitting the time
averaged MSD traces into two distinct populations, see Figs.~\ref{fig-exp-tamsd}
and \ref{fig-exp-split}. The diffusion coefficient $D_\beta$ for the initial
part of the $\overline{\delta^2(\Delta)}$ traces is also split for intermediate
$x_0$, see Fig.~\ref{fig-exp-D-split}. Relatively large $\overline{\delta^2}$
amplitudes with a $\Delta^{1/2}$ scaling emerge due to fast excursions into the
left semiaxis with large values of $D_\text{exp}(x)$. For larger $x_0$ the
fraction of $\overline{\delta^2}\sim\Delta^1$ traces increases. Around $x_0$=5 the
population splitting of temporal MSD is most prominent. Due to the presence
of small-amplitude $\overline{\delta^2}$ traces linear in $\Delta$, the mean
$\langle\overline{ \delta^2}\rangle$ in the simulations is slightly lower than
the theoretical $\Delta^{1/2}$-asymptote (\ref{exp-simple-TAMSD}). Note that
smaller diffusivity magnitudes $A$ have a similar effect as a larger $x_0$,
namely, the value of the exponent $\beta$ tends to change from 1/2 to 1 as $A$
decreases (not shown). 

\begin{figure}
\includegraphics[width=8cm]{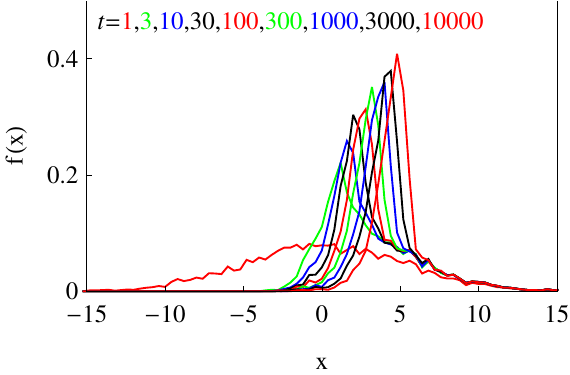}
\caption{Focusing and propagation of a diffusing front for a packet of particles
with initial normal distribution (\ref{profile-shape}).
We chose the parameters $w=5$ and $T=10^4$, and the results are averaged over
$N=5000$ traces. The diffusion time $t$ is indicated in the graph.}
\label{fig-exp-packet}
\end{figure}

This dramatic effect of the initial position $x_0$ affects the spreading of
a packet of particles diffusing in such a medium as well as the propagation of
diffusion fronts. Walkers, that are initially distributed normally according to
\begin{equation}
\label{profile-shape}
f(x)=\frac{1}{\sqrt{2\pi w^2}}\exp\left(-\frac{x^2}{2w^2}\right),
\end{equation}
escape the region of fast diffusivity after relatively few simulations steps,
due to the occurrence of relatively long jumps, that is, we observe a
superdiffusive front propagation. Because of this, a peak in the normalised
profiles develops at $x>0$, resembling the peak in the PDF,
Fig.~\ref{fig-exp-packet}. Slow particles
starting at $x_0\gg 1$ remain trapped in the slow-diffusion region for
long times, corresponding to the nearly unaltered right wing of the
distribution. Later on, the diffusion front exhibits a slow propagation
reminiscent of the slow MSD scaling (\ref{MSD-exp-theory}). Clearly, the
traces initiated at different $x_0$ values will have different ergodic
characteristics and the ergodic properties of the packet of diffusive
particles will change upon the spatial spreading. 

\begin{figure}
\includegraphics[width=7cm]{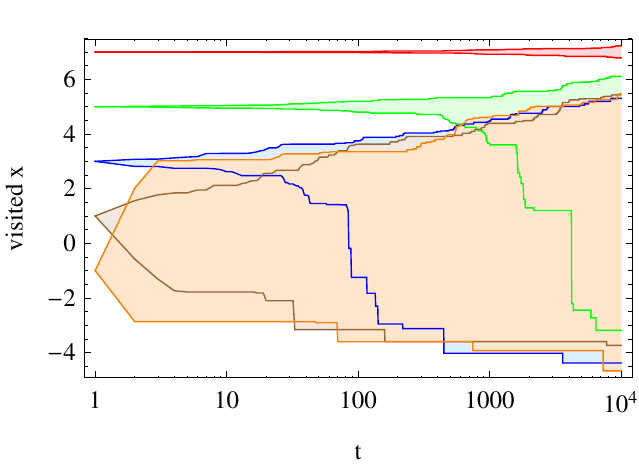}
\caption{Space exploration by particles diffusing in a medium with exponentially
distributed diffusion coefficient, shown for various initial conditions, $x_0=7$,
5, 3, 1, and $-1$. The trajectories are of length $T=10^{4}$.}
\label{fig-exp-explore}
\end{figure}

One final dynamic characteristic of HDPs is the exploration of space. This
property is relevant, for instance, for the random localisation of `targets'
by diffusing particles. The first-passage dynamics to such a target will be
strongly affected by the target position in our strongly non-homogeneous 
scenario for the exponentially distributed diffusivities. The results of our
simulations show that for large initial particle positions, $x_0\gg1$, the space
exploration is nearly symmetric and the diffusivity is small (small spread
around the initial position $x_0$). For moderate $x_0>0$ excursions into the
high diffusivity left semi-axis occur more frequently and earlier during the
time evolution, as underlined in Fig.~\ref{fig-exp-explore}. For $x_0=0$ the
exploration of both half-spaces is nearly equally fast. For negative $x_0$ with
large modulus the particles quickly escape from the region of high diffusivity
and the positive half-space is explored faster. The boundary of this exploration
front in the positive semi-axis appears to approach a universal curve for $x_0
\lesssim3$.

\section{Logarithmically varying diffusivity}
\label{sec-log}

To complete our analysis of diffusion processes with spatially varying
diffusion coefficients, to contrast the previous cases of power-law and
exponential variation, we now turn to the case slowly varying diffusivity.
More concretely we study the HDP process with logarithmic $x$ dependence
(\ref{D-log}) of the diffusion coefficient and perform a similar analysis as
pursued in the previous two Sections.

\subsection{PDF and ensemble averaged MSD}

\begin{figure}
\includegraphics[width=8cm]{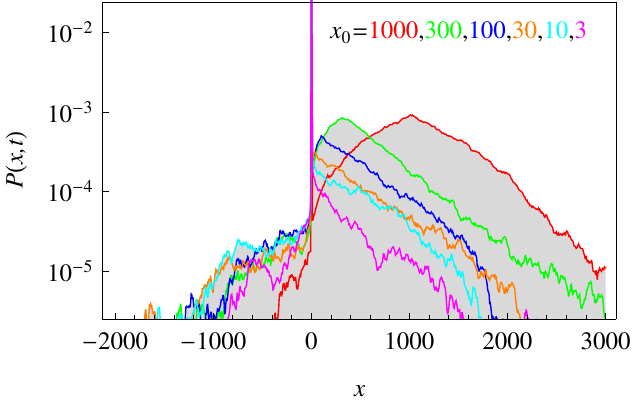}
\caption{PDF of the HDP with logarithmically varying diffusion coefficient,
computed for different initial conditions. The parameters are $T=10^5$, $A=1$,
and $\bar x=1$.}
\label{fig-log-pdf-x0}
\end{figure}

\begin{figure}
\includegraphics[width=8cm]{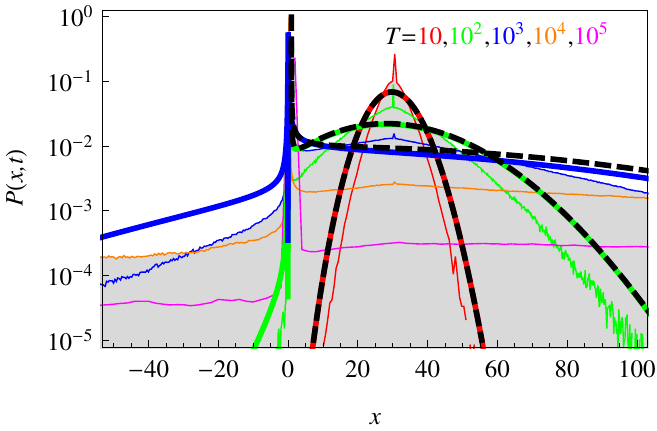}
\caption{PDF of the HDP with logarithmic space dependence of the diffusivity,
obtained from simulations with different $T$. The dashed black curves represent
Eq.~(\ref{log-pdf}), the coloured curves correspond to numerical solution of
Eq.~(\ref{diff-equation-strat}). The two sets of curves agree well for short
$T$. Parameters: $x_0=30$ and $A=1$.}
\label{fig-log-pdf}
\end{figure}

Using the same change of variables for the concrete form (\ref{log-diff-approx})
of $D(x)$ we find that ($\overline{x}=1$)
\begin{eqnarray}
\nonumber
y(x)&=&\int_{x_0}^x\frac{dx'}{\sqrt{2D_{\text{log}}(x')}}\\
&&\hspace*{-1.2cm}
=\frac{2}{A}\left[x(t)F\left(\sqrt{\log x}\right)-x_0F\left(\sqrt{\log x_0}
\right)\right],
\label{var-change-log}
\end{eqnarray}
where, we introduce Dawson's integral
\begin{equation}
F(z)=e^{-z^2}\int_0^ze^{y^2}dy.
\end{equation}
The PDF obtained from the PDF (\ref{wiener-pdf}) of the Wiener process then
assumes the form 
\begin{eqnarray}
\nonumber
P(x,t;x_0)&=&\frac{1}{\sqrt{2\pi tA^2\log x}}\\
&&\hspace*{-2.0cm}
\times\exp\left(-\frac{2\left[{xF\left(\sqrt{\log x}\right)-x_0F\left(\sqrt{
\log x_0}\right)}\right]^2}{A^2t}\right).
\label{log-pdf}
\end{eqnarray}
We simulated discretised HDPs with logarithmically varying diffusion coefficient,
Eq. (\ref{D-log}). This process features a region of low diffusivity around the
origin $x=0$. This region tends to trap particles diffusing in from higher
diffusivity regions,
and particles initially positioned close to the origin will escape this
region only very slowly. The PDF thus features two maxima, as shown in
Fig.~\ref{fig-log-pdf-x0}. The first maximum is due to the initial particle
position at $x=x_0$, while the second one at $x=0$ represents particles in
the low diffusivity zone around the origin. The initial spreading can be
captured by a shifted Gaussian bell curve with a renormalised diffusivity. For
longer trajectories the particles accumulate progressively at $x=0$ and the PDF
develops a tail at $x\gg x_0$, compare also Fig.~\ref{fig-log-pdf}.

These features can be quantitatively understood from the analytical shape
shape (\ref{log-pdf}) of the PDF. With increasing $x_0$, the gradient of the
diffusivity $D_{\log}(x)$ on the length scale covered by the diffusing particle
decreases and the HDP approaches regular Brownian motion, see also below and
in Fig.~\ref{fig-log-tamsd}. The trapping effect at $x=0$ becomes amplified for
larger magnitudes of $A$ (not shown).

Direct numerical solution of
Eq.~(\ref{diff-equation-strat}) for the logarithmic form of the diffusion
coefficient, Eq.~(\ref{D-log}), was obtained for moderate lengths $T$ of the
time series \cite{note-num-sol}. Eq.~(\ref{log-pdf}) describes the numerical
results quite well and also agrees well with the results of our stochastic
simulations, as shown in Fig.~\ref{fig-log-pdf}. 

Numerical integration of the analytical expression (\ref{log-pdf}) shows that
the particle's ensemble averaged MSD follows the linear Brownian time dependence,
with a renormalised diffusivity and the initial value $x_0^2$,
\begin{equation}
\label{MSD-log-theory}
\left<x^{2}(t)\right>\approx x_0^2+2D_\text{log}(x_0)t.
\end{equation}
This finding is is in good agreement with our stochastic simulations, see the
black dashed curves in Fig.~\ref{fig-log-tamsd}.

\subsection{Time averaged MSD, amplitude scatter, and ergodicity breaking }

\begin{figure}
\includegraphics[width=9cm]{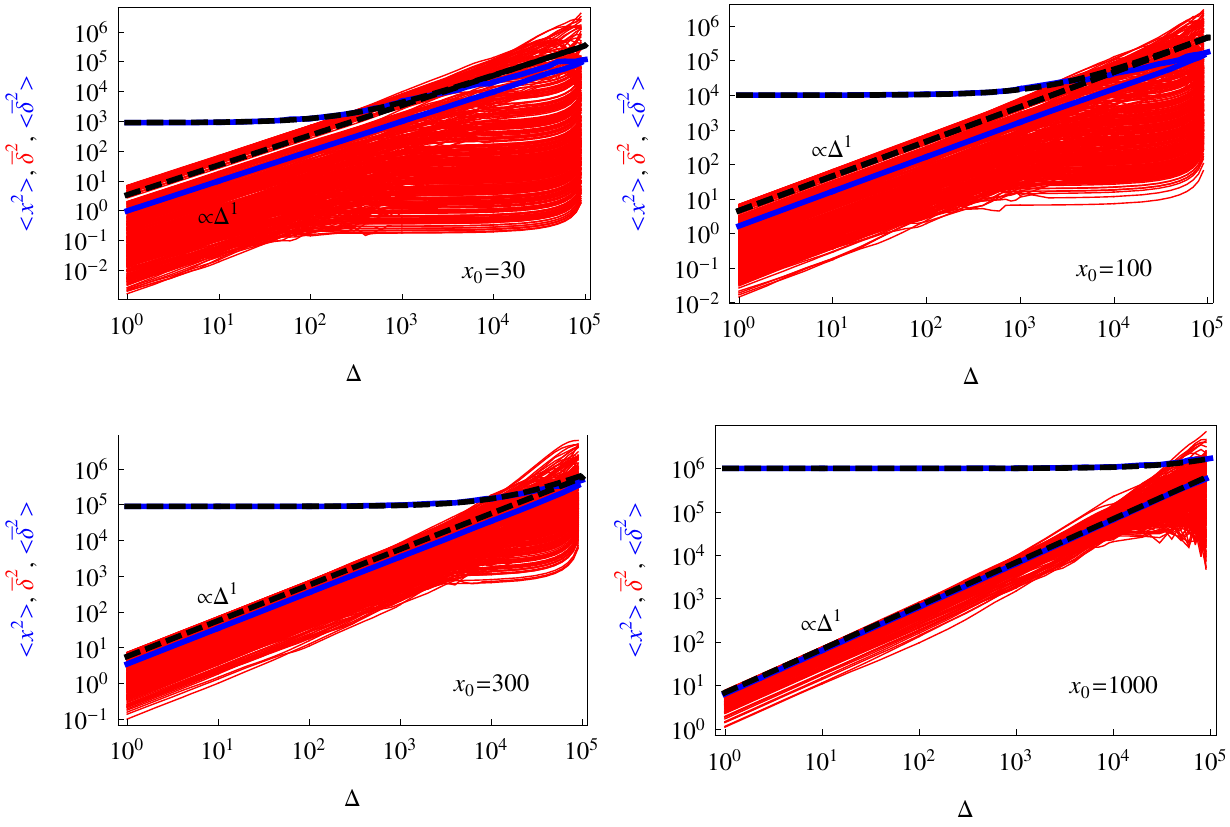}
\caption{Ensemble and time averaged MSDs of individual trajectories for the HDP
with logarithmically varying diffusion coefficient. The black dashed lines
represent Eq.~(\ref{MSD-log-theory}) for $\langle x^2\rangle$ and
Eq.~(\ref{log-tamsd}) for $\langle\overline{\delta^2}\rangle$. The parameters
are the same as in Fig.~\ref{fig-log-pdf-x0}.}
\label{fig-log-tamsd}
\end{figure}

The particle displacement $x(y)$ for the logarithmic dependence of the diffusion
coefficient is a non-trivial function of the Wiener process $y(t)$, as
demonstrated by Eq.~(\ref{var-change-log}), and it is hard to get a general
expression for $\overline{\delta^2(\Delta)}$. In the short time limit $t\to0$,
however, expanding Dawson's integral for $|x-x_0|\ll1$, one finds a linear
relation of $x(y)$, namely,
\begin{equation}
x(t)\approx x_0+\log^{1/2}[x_0]Ay(t).
\label{log-x-y}
\end{equation}
This relation resembles Eq.~(\ref{exp-x-via-y-large-x0}) for the case of
exponentially varying diffusivity. Then, using Eq.~(\ref{exp-pi-y}), the
position correlations become
\begin{equation}
\left\langle x(t)x(t+\Delta)\right\rangle=\left\langle x^2(t)\right\rangle
\approx x_0^2+\log[x_0]A^2\Delta.
\label{log-x-corr}
\end{equation}
In this limit, the time averaged MSD is a linear function of the lag time
$\Delta$ with an effective diffusivity depending on the initial particle
position,
\begin{equation}
\left<\overline{\delta^2(\Delta)}\right>\approx2D_\text{log}(x_0)\Delta.
\label{log-tamsd}
\end{equation}
This linear scaling is identical with the result Eq.~(\ref{tamsd-exp-effective-d})
for exponentially varying diffusivity. It is also in agreement with computer
simulations for $x_0\gg 1$, as shown in Fig.~\ref{fig-log-tamsd}. In this
regime, the ergodicity breaking parameter vanishes in the limit
$\Delta/T\to0$.

In contrast to the initial plateau $x_0^2$ of the ensemble averaged MSD
(\ref{MSD-log-theory}), the time averaged MSD $\overline{\delta^2}$ starts
linearly in the lag time $\Delta$ and in fact stays linear for those particles,
that do not become trapped. The particles that eventually do become trapped
in the low-diffusivity zone give rise to a stalling of the time averaged MSD
$\overline{\delta^2}$ so that we observe a population splitting between mobile
and immobile fractions with local scaling exponents $\beta\approx1$ and $\beta
\approx0$, respectively. Trapping is obviously strongest for small $x_0$, for
which the spread of the temporal MSD is also the largest. Due to these immobile
particles, the analytical value (\ref{log-tamsd}) is higher than the actual value
$\langle\overline{\delta^2}\rangle$ from the simulations
(Fig.~\ref{fig-log-tamsd}). Such particle
immobilisation and its effect on  $\langle\overline{\delta^2}\rangle$
is similar to that observed for continuous time random walks with ageing
\cite{johannes}, see discussion in Sec.~\ref{sec-out}.

For more remote initial positions, $x_0\gg1$ the `diffusion trap' at $x=0$ is
not strong enough, the fraction of normal traces $\overline{\delta^2}\sim\Delta$
grows, and $\langle\overline{\delta^2}\rangle$ is nicely described by
Eq.~(\ref{log-tamsd}), compare the dashed black line in Fig.~\ref{fig-log-tamsd}.
At smaller $A$, the trapping propensity of the trap is impeded (not shown). 

\begin{figure}
\includegraphics[height=9cm,angle=270]{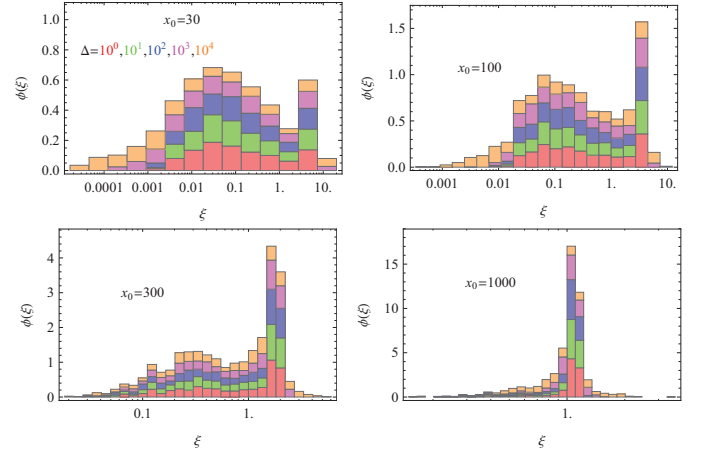}
\caption{Amplitude scatter of the time averaged MSD $\overline{\delta^2}$ for
different initial positions $x_0$ for the HDP with logarithmic varying
diffusion coefficient. In this plot, bins of stacked histograms do not overlap.
The fraction of traces with small magnitudes of $\overline{\delta^2}$ grows as
$x_0$ decreases. The parameters are the same as in Fig.~\ref{fig-log-pdf-x0}.}
\label{fig-phi-xi-log}
\end{figure}

\begin{figure}
\includegraphics[width=7cm]{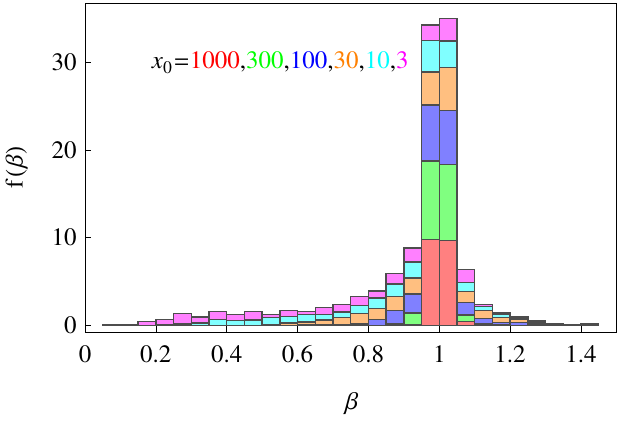}
\caption{Apparent scaling exponent of the initial behaviour of the time averaged
MSD $\overline{\delta^2}$ revealing a greater spread at small $x_0$, for which
the fraction of trapped trajectories is larger. The parameters are the same as
in Fig.~\ref{fig-log-pdf-x0}.}
\label{fig-log-split}
\end{figure}

The amplitude scatter distribution of individual traces $\overline{\delta^2}$
is broad for small values of the initial position $x_0$, as demonstrated in
Fig.~\ref{fig-phi-xi-log}. The distribution in fact also exhibits a certain
bi-modality due to the population splitting into mobile and immobile particles.
For longer lag times $\Delta$ the fraction of trapped particles increases and
the peak of the scatter distribution around $\overline{\delta^2(\Delta)}=0$
becomes more pronounced. The local scaling exponent $\beta$, however, is
predominantly unimodal and centred around unity for larger values of $x_0$,
compare the histograms in Fig.~\ref{fig-log-split}.

\begin{figure}
\includegraphics[width=8cm]{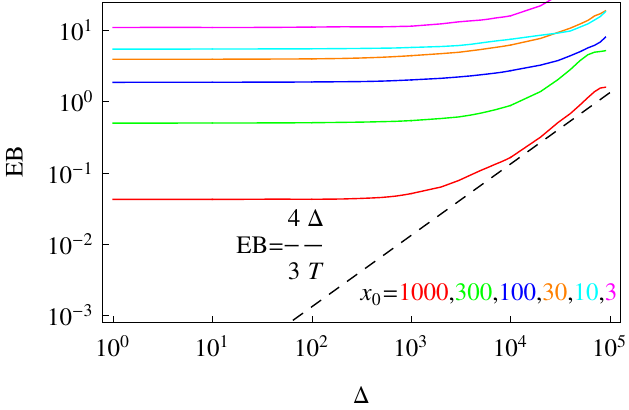}
\caption{Ergodicity breaking parameter EB for logarithmically varying diffusion
coefficient. It approaches the Brownian behaviour (\ref{eb-bm}) for large values
of the initial position $x_0$. The parameters are the same as in
Fig.~\ref{fig-log-pdf-x0}.}
\label{fig-log-eb}
\end{figure}

Fig.~\ref{fig-log-eb} illustrates the non-ergodic nature of HDPs with logarithmic
$x$-dependence of the diffusivity. We observe that for small modulus of the
initial position $x_0$ a substantial fraction of particles is trapped at $x$=0
and the ergodicity breaking parameter (\ref{EB1}) is relatively large, namely,
$\mathrm{EB}\gg1$ \cite{eb1-vs-eb2}. For $x_0\gg1$ the HDP is nearly ergodic,
recovering the self-averaging property of normal diffusion. Note that as the
length $T$ of the time series grows, the HDP approaches the ergodic behaviour at
considerably larger $x_0$ values, compare Fig.~\ref{fig-log-eb} as well as
Fig.~A2 in the Appendix.

\begin{figure}
\includegraphics[width=8cm]{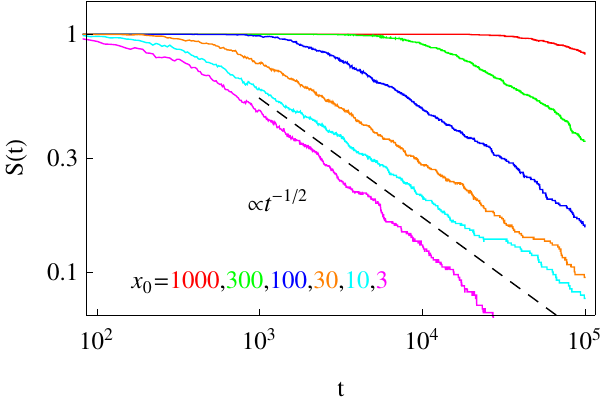}
\caption{Survival probability of non-trapped particles for $D=D_\text{log}(x)$.
Starting positions $x_0$ are indicated. The length of the trajectories is $T=
10^5$. Walkers with an amplitude smaller than the cutoff $x_\star=10^{-5}$
are considered trapped.}
\label{fig-log-surv}
\end{figure}

We conclude this Section with the analysis of the survival probability $S(t)$,
that measures the fraction of particles remaining mobile as function of the
diffusion time $t$. This survival probability is thus a dynamic characteristic
for the immobilisation of particles over time in the trapping potential effected
by the form (\ref{D-log}) of the diffusion coefficient. As discussed above, at
small initial distances $x_0$ from the capturing well, the fraction of stalled
walkers grows. For larger $x_0$, a larger fraction of particles remain mobile,
corresponding to a larger value of $S(t)$ at the same time $t$. This behaviour
is shown in Fig.~\ref{fig-log-surv}. Computer simulations show that, independent
of the starting position, the survival probability decreases for long times as
\begin{equation}
S(t)\sim\frac{1}{\sqrt{t}},
\end{equation}
see the dashed line in Fig.~\ref{fig-log-surv} representing the inverse square
root scaling. For particles starting at larger $x_0$ the onset of this scaling
is naturally delayed to longer times $t$.

\section{Conclusions and Outlook}
\label{sec-out}

We analysed a model for HDPs with distance-dependent diffusivities that exhibit
sub-, super-, and ultra-slow diffusion as well as weak ergodicity breaking.
Power-law, exponential and logarithmic variations of the diffusion coefficient
were examined. This framework can be applied to other variants of the spatial
dependence $D(x)$ of the diffusion coefficient \cite{note-stretched-exp}.
Our results may find applications in a wide variety of spatially heterogeneous
media. A particular example is the viral
infection dynamics, as a mathematical rational to discriminate nearly Brownian
and anomalous populations of diffusing viral particles, which was observed by
single particle tracking in living bacteria  \cite{seisen01}. For this purpose,
an extension of the analytical and computational schemes for HDPs in higher
dimensions is currently in progress \cite{inprep}.

In particular, for an exponentially varying diffusivity we showed that the
initial condition of the system have a vital impact on the time dependence of
the process. Specifically, depending on the gradient of the particle diffusivity
over the first steps of a trajectory, the scaling of the temporal MSD may become
anomalous [$\overline{\delta^2(\Delta)}\simeq\Delta^{1/2}$] and thus lead to a
population splitting compared with the traces with linear scaling
[$\overline{\delta^2(\Delta)}\simeq\Delta$]. The time averaged traces with
this anomalous scaling $\overline{\delta^2(\Delta)}\sim\Delta^{1/2}$ progressively
drive the system toward stronger deviations from ergodicity. We also examined the
asymmetry
in the spatial exploration patterns, which will affect the efficiency of
diffusion limited processes in such a medium.

For the case of a logarithmically varying diffusion coefficient with an
associated trap of vanishing diffusivity at the origin, we also observed weakly
non-ergodic behaviour with split populations with respect to the time averaged
MSD $\overline{\delta^2}$. Here, stalled traces with $\overline{\delta^2(\Delta)}
\sim\mathrm{const.}$ separate from mobile ones. For particles starting far from
the trap at the origin, however, the ensemble and time averaged characteristics
can be captured in terms of a Brownian-style motion with renormalised diffusivity
$D(x_0)$. 

Let us contrast these observations with the results of the subdiffusive continuous
time random walk model, compare Ref.~\cite{pccp}. Due to the underlying long
tailed distribution of trapping times $\tau$, $\psi(\tau)\sim\tau^{-(1+\alpha)}$
with $0<\alpha<1$, the characteristic waiting time $\langle\tau\rangle$ for this
system diverges. The ergodicity breaking then occurs naturally because the
lack of a finite microscopic time scale $\langle\tau\rangle$ negates the
existence of a long measurement time $T$ limit and thus the system remains
non-stationary. For the HDPs considered here, the violation of ergodicity is
solely due to the spatial variation of the diffusion process, and the anomalous
diffusion is due to the multiplicative nature of the noise.

Regarding the population splitting in terms of the time averaged MSD $\overline{
\delta^2}$, we note that a similar effect was recently analysed for continuous
time random walks in the presence of strong ageing \cite{johannes}. In that
case the proportion of immobile versus trapped walkers was shown to grow with
the age $t_a$ of the process. Concurrently, the ergodicity breaking parameter
for such strong ageing diverges for $t_a\gg T$ in the form $\mathrm{EB}\sim(t_a/T)
^{-(1+\alpha)}$, while for the population of exclusively mobile particles one
finds $0<\text{EB}_{m}\leq1$ in the same limit \cite{johannes}. For HDPs with
exponential variation of the diffusivity we similarly observe that the trapped
particles contribute large values to the $\mathrm{EB}$ parameter, while slowly
but normally diffusing particles far from the trap remain nearly ergodic. 

\begin{figure*}
\includegraphics[width=6cm]{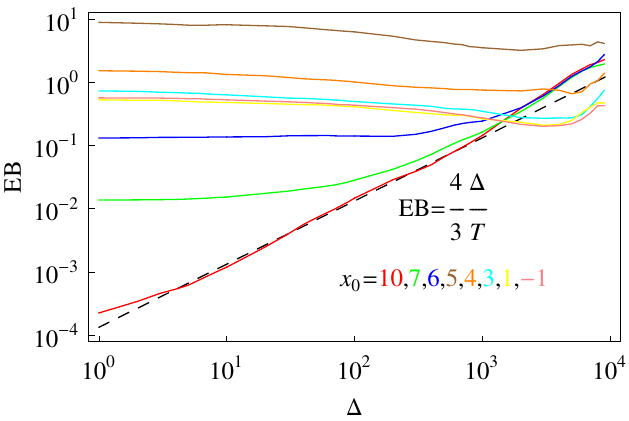}
\includegraphics[width=6cm]{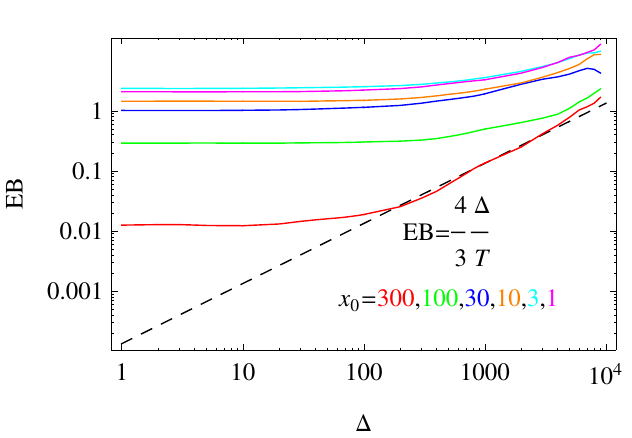}
\caption{Left: Ergodicity breaking parameter for $D=D_\text{exp}(x)$ and
Right: Ergodicity breaking parameter for $D=D_\text{log}(x)$, both for
$T=10^4$.}
\label{fig-exp-eb-10in4}
\label{fig-log-eb-10in4}
\end{figure*}

Experimentally, a coexistence of ergodic and non-ergodic diffusion pathways
was observed for the motion of ion channels in plasma membranes and of insulin
granules in the cytosol of living cells \cite{weigel}. Similarly, direct
tracking of proteins and cajal bodies diffusing in the cell nucleus revealed
the existence of two particle populations with distinct mobilities
\cite{kubi01,platani}. Strongly restricted diffusion in the crowded nucleus
environment, with normal and anomalous components possibly occurring on
different length- and time-scales, may produce such a separation effect.

\acknowledgments

The authors thank E. Barkai, A. Chechkin, A. Godec, and I. Goychuk for
stimulating discussions. Funding from the Academy of Finland (FiDiPro scheme,
RM) and the German Research Council (DFG Grant CH 707/5-1, AGC) is acknowledged. 
RM thanks the Mathematical Institute of the University of Oxford for financial
support as an OCCAM Visiting Fellow.

\begin{appendix}

\section{Ergodicity breaking parameter for shorter trajectories}

To illustrate the approach to ergodicity, we present graphs for the ergodicity
breaking parameter for trajectories, that are 10 times shorter than those used
in the the majority of Figures in the main text. These Figures are referenced
in the main text.

\end{appendix}


\begin{thebibliography}{99}

\bibitem{bouchaud} J.-P. Bouchaud and A. Georges, Phys. Rep. \textbf{195}, 127
(1990).

\bibitem{report} R. Metzler and J. Klafter, Phys. Rep. \textbf{339}, 1
(2000); J. Phys. A \textbf{37}, R161 (2004).

\bibitem{scher} H. Scher and E. W. Montroll, Phys. Rev. B \textbf{12}, 2455
(1975).

\bibitem{amblard} F. Amblard, A. C. Maggs, B. Yurke, A. N. Pargellis, and S.
Leibler, Phys. Rev. Lett. \textbf{77}, 4470 (1996).

\bibitem{weitz} I. Y. Wong, M. L. Gardel, D. R. Reichman, E. R. Weeks, M. T.
Valentine, A. R. Bausch, and D. A. Weitz, Phys. Rev. Lett. \textbf{92}, 178101
(2004).

\bibitem{chaikin} Q. Xu, L. Feng, R. Sha, N. C. Seeman, and P. M. Chaikin,
Phys. Rev. Lett. \textbf{106}, 228102 (2011).

\bibitem{harvey} H. Scher, G. Margolin, R. Metzler, J. Klafter, and B. Berkowitz,
Geophys. Res. Lett. \textbf{29}, 1061 (2002).

\bibitem{swinney} T. H. Solomon, E. R. Weeks, and H. L. Swinney, Phys. Rev.
Lett. \textbf{71}, 3975 (1993).

\bibitem{stapf} S. Stapf, R. Kimmich, and R.-O. Seitter, Phys. Rev. Lett.
\textbf{75}, 2855 (1995).

\bibitem{ott} A. Ott, J.-P. Bouchaud, D. Langevin, and W. Urbakh, Phys. Rev.
Lett. \textbf{65}, 2201 (1990).

\bibitem{pt} E. Barkai, Y. Garini, and R. Metzler, Phys. Today \textbf{65},
 29 (2012).

\bibitem{saxton} M. J. Saxton, Biophys. J. \textbf{72}, 1744 (1997);
M. J. Saxton and K. Jacobson, Annu. Rev. Biophys. Biomol. Struct. {\bf 26},
373 (1997).

\bibitem{franosch13} F. H{\"o}fling and T. Franosch, Rep. Prog. Phys. \textbf{76},
046602 (2013).

\bibitem{bress13} P. C. Bressloff and J. M. Newby, Rev. Mod. Phys. \textbf{85},
135 (2013).

\bibitem{micro} S. Yamada, D. Wirtz, and S. C. Kuo, Biophys. J. \textbf{78},
1736 (2000).

\bibitem{lene} J.-H. Jeon, V. Tejedor, S. Burov, E. Barkai, C. Selhuber-Unkel, K.
Berg-S{\o}rensen, L. Oddershede, and R. Metzler, Phys. Rev. Lett. \textbf{106},
048103 (2011).

\bibitem{tabei} S. M. A. Tabei, S. Burov, H. Y. Kim, A. Kuznetsov, T. Huynh,
J. Jureller, L. H. Philipson, A. R. Dinner, and N. F. Scherer, Proc. Natl.
Acad. Sci. USA \textbf{110}, 4911 (2013).

\bibitem{taylor} M. A. Taylor, J. Janousek, V. Daria, J. Knittel, B. Hage,
H.-A. Bachor, and W. P. Bowen, Nature Photonics \textbf{7,} 229 (2013).

\bibitem{golding} I. Golding and E. C. Cox, Phys. Rev. Lett. \textbf{96}
 098102 (2006).

\bibitem{weber} S. C. Weber, A. J. Spakowitz, and J. A. Theriot, Phys. Rev.
Lett. \textbf{104}, 238102 (2010).

\bibitem{bronstein} I. Bronstein, Y. Israel, E. Kepten, S. Mai, Y. Shav-Tal,
E. Barkai, and Y. Garini, Phys. Rev. Lett. \textbf{103}, 018102 (2009).

\bibitem{guigas} G. Guigas, C. Kalla, and M. Weiss, Biophys. J. \textbf{93},
316 (2007).

\bibitem{caspi} A. Caspi, R. Granek and M. Elbaum, Phys. Rev. Lett. \textbf{85},
5655 (2000); Phys. Rev. E \textbf{66}, 011916 (2002).

\bibitem{seisen01} G. Seisenberger, M. U. Ried, T. Endre{\ss}, H. B{\"u}ning,
M. Hallek, and C. Br{\"a}uchle, Science \textbf{294}, 1929 (2001).

\bibitem{brauchle02} C. Brauchle, G. Seisenberger, T. Endre{\ss}, M. U. Ried,
H. B{\"u}ning, and M. Hallek, Chem. Phys. Chem. \textbf{3}, 299 (2002).

\bibitem{bruno09} L. Bruno, V. Levi, M. Brunstein, M. A. Desposito,
Phys. Rev. E 80, 011912 (2009).

\bibitem{fradin05} D. S. Banks and C. Fradin, Biophys. J. \textbf{89}, 2960
(2005).

\bibitem{weigel} A. V. Weigel, B. Simon, M. M. Tamkun, and D. Krapf,
Proc. Nat. Acad. Sci. USA \textbf{108}, 6438 (2011).

\bibitem{weiss03} M. Weiss, H. Hashimoto, and T. Nilsson, Biophys. J.
\textbf{84}, 4043 (2003).

\bibitem{jeon-lipids} J.-H. Jeon, H. Martinez-Seara Monne, M. Javanainen, and
R. Metzler, Phys. Rev. Lett. \textbf{109}, 188103 (2012); M. Javanainen, H.
Hammaren, L. Monticelli, J.-H. Jeon, R. Metzler, and I. Vattulainen,
Faraday Disc. \textbf{161}, 397 (2013).

\bibitem{kneller} G. R. Kneller, K. Baczynski, and M. Pasenkiewicz-Gierula,
J. Chem. Phys. \textbf{135}, 141105 (2011).

\bibitem{akimoto} T. Akimoto, E. Yamamoto, K. Yasuoka, Y. Hirano, and M. Yasui,
Phys. Rev. Lett. \textbf{107}, 178103 (2011).

\bibitem{robert} D. Robert, T.-H. Nguyen, F. Gallet, and C. Wilhelm,
PLoS ONE \textbf{4}, e10046 (2010).

\bibitem{igor} I. M. Sokolov, Soft Matter \textbf{8}, 9043 (2012).

\bibitem{pccp} S. Burov, J.-H. Jeon, R. Metzler, and E. Barkai,
Phys. Chem. Chem. Phys. \textbf{13}, 1800 (2011).

\bibitem{goychuk} I. Goychuk, Phys. Rev. E \textbf{80}, 046125 (2009);
Adv. Chem. Phys. \textbf{150}, 187 (2012).

\bibitem{montroll} E. W. Montroll and G. H. Weiss, J. Math. Phys. \textbf{6},
167 (1965).

\bibitem{ffpe} R. Metzler, E. Barkai, and J. Klafter, Phys. Rev. Lett.
\textbf{82}, 3563 (1999).

\bibitem{nctrw} J.-H. Jeon, E. Barkai, and R. Metzler, J. Chem. Phys.
(at press).

\bibitem{mandelbrot} B. B. Mandelbrot and J. W. van Ness, SIAM Rev. \textbf{1},
42 (1968); A. N. Kolmogorov, Dokl. Acad. Sci. USSR \textbf{26}, 115 (1940).

\bibitem{lutz} E. Lutz, Phys. Rev. E \textbf{64}, 051106 (2001).

\bibitem{weiss} D. Ernst, M. Hellmann, J. K{\"o}hler, and M. Weiss, Soft
Matter \textbf{8}, 4886 (2012).

\bibitem{fulinski} A. Fuli{\'n}ski, Phys. Rev. E \textbf{83}, 061140 (2011);
J. Chem. Phys. \textbf{138}, 021101 (2013).

\bibitem{havlin} S. Havlin and D. Ben-Avraham, Adv. Phys. \textbf{36}, 695
(1987).

\bibitem{klemm} A. Klemm, R. Metzler, and R. Kimmich, Phys. Rev. E \textbf{65},
021112 (2002).

\bibitem{wong} M. F. Shlesinger, J. Klafter, and Y. M. Wong, J. Stat.
Phys. \textbf{27}, 499 (1982)

\bibitem{lw1} M. Niemann, H. Kantz, and E. Barkai, Phys. Rev. Lett.
\textbf{110}, 140603 (2013).

\bibitem{zukla} G. Zumofen and J. Klafter, Phys. Rev. E \textbf{51}, 1818 (1995);
\emph{ibid.} \textbf{47}, 851 (1993).

\bibitem{aljaz} A. Godec and R. Metzler, Phys. Rev. Lett. \textbf{110}, 020603
(2013); Phys. Rev. E \textbf{88}, 012116 (2013); D. Froemberg and E. Barkai,
Phys. Rev. E \textbf{87}, 030104(R) (2013); E-print arXiv:1306.2036.

\bibitem{elfDX11} B. P. English, V. Hauryliuk, A. Sanamrad, S. Tankov, N. H.
Dekker, and J. Elf, Proc. Natl. Acad. Sci. \textbf{108}, E365 (2011).

\bibitem{langDX11} T. K{\"u}hn, T. O. Ihalainen, J. Hyv{\"a}luoma, N. Dross,
S. F. Willman, J. Langowski, M. Vihinen-Ranta, and J. Timonen, PLoS One
\textbf{6}, e22962 (2011).

\bibitem{platani} M. Platani, I. Goldberg, A. I. Lamond, and J. R. Swedlow,
Nature Cell Biol. \textbf{4}, 502 (2002).

\bibitem{hagger95} R. Haggerty and S. M. Gorelick, Water Res. Res. \textbf{31},
 2383 (1995).

\bibitem{hdp-ctrw} M. Dentz, P. Gouze, A. Russian, J. Dweik, and F. Delay, Adv.
Water Res. \textbf{49}, 13 (2012).

\bibitem{srokowski06} T. Srokowski and A. Kaminska, Phys. Rev. E \textbf{74},
021103 (2006).

\bibitem{srokowski08} T. Srokowski, Phys. Rev. E \textbf{78}, 031135 (2008).
 
\bibitem{silva11} A. T. Silva, E. K. Lenzi, L. R. Evangelista, M. K. Lenzi,
H. V. Ribero, and A. A. Tateishi, J. Math. Phys., \textbf{52} 083301 (2011).

\bibitem{chechkin-inhom} A. V. Chechkin, R. Gorenflo and I. M. Sokolov,
J. Phys. A: Math. Gen. \textbf{38}, L679  (2005).

\bibitem{maglom} L. F. Richardson, Proc. Roy. Soc. London, Ser. A \textbf{110},
709 (1926); A. S. Monin and A. M. Yaglom, Statistical Fluid Mechanics
(MIT Press, Cambdridge MA, 1971).

\bibitem{fractals-proc} B. O'Shaughnessy and I. Procaccia, Phys. Rev. Lett.
\textbf{54}, 455 (1985).

\bibitem{yazmin} Y. Meroz, I. Eliazar, and J. Klafter, J. Phys. A \textbf{42},
434012 (2009); Y. Meroz, I. M. Sokolov, and J. Klafter, Phys. Rev. Lett.
\textbf{110}, 090601 (2013).

\bibitem{hdp13} A. G. Cherstvy, A. V. Chechkin, and R. Metzler, 
http://arxiv.org/abs/1303.5533

\bibitem{web} J.-P. Bouchaud, J. Phys. I (Paris) \textbf{2}, 1705 (1992);
G. Bel and E. Barkai, Phys. Rev. Lett. \textbf{94}, 240602;
A. Rebenshtok and E. Barkai, \emph{ibid.} \textbf{99}, 210601 (2007);
M. A. Lomholt, I. M. Zaid, and R. Metzler, Phys. Rev. Lett.
\textbf{98}, 200603 (2007).

\bibitem{he} Y. He, S. Burov, R. Metzler and E. Barkai,
Phys. Rev. Lett. \textbf{101}, 058101 (2008).

\bibitem{ariel} A. Lubelski, I. M. Sokolov, and J. Klafter, Phys. Rev. Lett.
\textbf{100}, 250602 (2008).

\bibitem{pnas} S. Burov, R. Metzler, and E. Barkai, Proc. Natl. Acad. Sci. USA
\textbf{107}, 13228 (2010).

\bibitem{johannes} J. H. P. Schulz, E. Barkai, and R. Metzler, Phys. Rev. Lett.
\textbf{110}, 020602 (2013); E. Barkai, \emph{ibid.} \textbf{90}, 104101 (2003).

\bibitem{vincent} M. Magdziarz, R. Metzler, W. Szczotka, and P. Zebrowski,
Phys. Rev. E \textbf{85}, 051103 (2012); V. Tejedor and R. Metzler, J.
Phys. A \textbf{43}, 082002 (2010).

\bibitem{lomholt} M. A. Lomholt, L. Lizana, R. Metzler, and T.
Ambj{\"o}rnsson, Phys. Rev. Lett. \textbf{110}, 208301 (2013).

\bibitem{deng} W. Deng and E. Barkai, Phys. Rev. E \textbf{79}, 011112
(2009).

\bibitem{jae} J.-H. Jeon and R. Metzler, Phys. Rev. E \textbf{85}, 021147
(2012); J.-H. Jeon, N. Leijnse, L. B. Oddershede, and R. Metzler,
New J. Phys. \textbf{15}, 045011 (2013).

\bibitem{olivier} S. Condamin, V. Tejedor, R. Voituriez, O. B{\'e}nichou, and
J. Klafter, Proc. Natl. Acad. Sci. USA \textbf{105}, 5675 (2008).

\bibitem{tejedor} V. Tejedor, O. B{\'e}nichou, R. Voituriez, R. Jungmann, F.
Simmel, C. Selhuber-Unkel, L. Oddershede, and R. Metzler, Biophys. J.
\textbf{98}, 1364 (2010).

\bibitem{p-var} M. Magdziarz, A. Weron, K. Burnecki, and J. Klafter, Phys.
Rev. Lett. \textbf{103}, 180602 (2009); M. Magdziarz and J. Klafter, Phys. Rev.
E \textbf{82}, 011129 (2010).

\bibitem{kepten} K. Burnecki, E. Kepten, J. Janczura, I. Bronshtein, Y. Garini,
and A. Weron, Biophys. J. \textbf{103}, 1839 (2012).

\bibitem{kevin} A. Robson, K. Burrage, and M. C. Leake, Trans. Roy. Soc. B
\textbf{368}, 20120029 (2013).

\bibitem{risken} H. Risken, The Fokker-Planck Equation, (Springer, Heidelberg,
1989).

\bibitem{rytov} S. M. Rytov, Introduction to Statistical Radio-Physics
(Defense Technical Information Center, Moscow, 1968).

\bibitem{nematodes} S. Hapca, J. W. Crawford, K. MacMillan, M. J. Wilson, and
L. M. Young, J. Theoret. Biol. \textbf{248}, 212 (2007); S. Hapca, J. W.
Crawford, and L. M. Young, J. Roy. Soc. Interfaces \textbf{6}, 111 (2009).

\bibitem{irradiate} J. Kowall, D. Peak, and J. W. Corbett, Phys. Rev. B
\textbf{13}, 477 (1976); A. G. Kesarev and V. V. Kondrat'ev, Phys. Metals
\& Metallogr. \textbf{108}, 30 (2009).

\bibitem{santos} S. B. Yuste, E. Abad, and K. Lindenberg, Phys. Rev. E
\textbf{82}, 061123 (2010).

\bibitem{srokowski09} T. Srokowski, Phys. Rev. E \textbf{79}, 040104 (2009).

\bibitem{katja} S. I. Denisov, S. B. Yuste, Y. S. Bystrik, H. Kantz, and K.
Lindenberg, Phys. Rev. E \textbf{84}, 061143 (2011);
S. I. Denisov, Yu. Bystrik, and H. Kantz, \emph{ibid.} \textbf{87},
022117 (2013);
J. Dr{\"a}ger and J. Klafter, Phys. Rev. Lett. \textbf{84}, 5998 (2000).

\bibitem{sinaj} Ya. G. Sinai, Theory Prob. Appl. \textbf{27}, 256 (1982).

\bibitem{doussal-sinai} P. Le Doussal et al., Phys. Rev. E \textbf{59}, 4795
(1999).

\bibitem{doussal98} L. Laloux and P. Le Doussal et al., Phys. Rev. E
\textbf{57}, 6296 (1998).

\bibitem{beta-note} To determine the exponent of
$\overline{\delta^2(\Delta)}\sim \Delta^\beta$ scaling at $\Delta\to 0$, we
first remove the last decade of temporal traces to improve the statistics. The
remaining  $T/10$ long trace is divided into two equal parts in the log-scale
for $\Delta$. Fitting the initial part of $\overline{\delta^2(\Delta)}$ with
log-sampled points by $\Delta^\beta$ gives the starting exponent
$\beta$. 

\bibitem{note-num-sol} The outcomes are almost insensitive to the choice
of boundary conditions in the simulation box and the PDF keeps its norm,
in contrast to the same computation scheme applied to the strongly asymmetric
$D_{\mathrm{exp}}(x)$ case. 

\bibitem{eb1-vs-eb2} Here we distinguish the two ergodicity breaking
parameters, Eqs. (\ref{EB1}) and (\ref{EB2}). The first one is the sufficient
condition of ergodicity, it involves only temporal moments and is  a robust
characteristics of the  process. The second one operates with 2nd moments
for ensemble and time averaged MSDs and therefore is a function of initial
conditions. From Fig. \ref{fig-log-eb} we observe e.g. that  $\mathcal{EB}\sim
1 $ at $x_0=1$, while the canonical $\mathrm{EB}$ parameter is far away from
its ergodic
value $\mathrm{EB}=0$. In contrast, at large $x_0$ we get that
$\mathcal{EB}
\ll 1$  because of large $\langle x^2\rangle$  starting at $t=0$ with $x_0^{2}$
and small $\langle\overline{\delta^2}\rangle$ amplitude. The EB parameter
follows however closely the Brownian law (\ref{eb-bm}) in a large region of
$\Delta$ .

\bibitem{note-stretched-exp} We have performed a similar type of analysis for
other exponentially varying forms of the diffusion coefficient, e.g., for a
stretched exponential $D(x)\sim\exp[-2\alpha|x|^{\gamma}]$. In the case of
$\gamma=2,$ for instance, the PDF exhibits two symmetric peaks. The scaling
of $\overline{\delta^2}$ traces depends on the particle starting position
$x_0$. For $x_0$ near a peak of PDF, the particle stays effectively trapped
that results in small $\overline{\delta^2}$ magnitudes. For the particles
jumping between the PDF peaks, the variation in position is large and so is the
$\overline{\delta^2}$  magnitude. The population splitting of temporal MSDs
takes place due to the existence of these two diffusion pathways. Initially,
we observe a linear growth $\langle\overline{\delta^2}\rangle\sim\Delta$,
while for the later stages of the trajectory a crossover to  $\langle
\overline{\delta^2}\rangle\sim \Delta^{1/2}$ scaling takes place.

\bibitem{inprep} A. G. Cherstvy, A. V. Chechkin, and R. Metzler (unpublished).

\bibitem{kubi01} T. Kues, R. Peters, and U. Kubitscheck,
Biophys. J. \textbf{80}, 2954 (2001).

\end{thebibliography}
\end{document}